\documentclass[12pt]{article} 
\usepackage{graphicx} 
\usepackage{amssymb} 
\usepackage{amsmath}

\begin{document} 
\begin{titlepage}
	\rightline{}
	
	
	\vskip 2cm 
	\begin{center}
		\large{{\bf A new mechanism for non-locality from string theory:\\
		UV-IR quantum entanglement and its imprints on the CMB}} 
	\end{center}
	
	\vskip 2cm 
	\begin{center}
		{Gregory Minton\footnote{\texttt{gminton@hmc.edu}}\ \ \ and\ \ \ Vatche Sahakian\footnote{\texttt{sahakian@hmc.edu}}}\\
	\end{center}
	\vskip 12pt \centerline{\sl Harvey Mudd College} \centerline{\sl Physics Department, 241 Platt Blvd.} \centerline{\sl Claremont CA 91711 USA}
	
	\vskip 2cm 
	\begin{abstract}
Puff field theories (PFT) arise as the decoupling limits of D3 branes in a Melvin universe and exhibit spatially non-local dynamics. Unlike other realizations of non-locality in string theory, PFTs have full SO(3) rotational symmetry. In this work, we analyze the strongly coupled regime of a PFT through gravitational holography. We find a novel mechanism at the heart of the phenomenon of non-locality: a quantum entanglement of UV and IR dynamics. In the holographic bulk, this translates to an apparent horizon splitting the space into two regions - with the UV completion of the PFT sitting at the horizon. We unravel this intricate UV-IR setting and devise a prescription for computing correlators that extends the original dictionary of holographic renormalization group. We then implement a cosmological scenario where PFT correlators set the initial conditions for primordial fluctuations. We compute the associated power spectrum of the CMB and find that the scenario allows for a distinct stringy signature.

	\end{abstract}
\end{titlepage}

\newpage \setcounter{page}{1} 
\section{Introduction and results}

Many recent areas at the forefront of fundamental physics access new realms of dynamics through non-local mechanisms. Such exotic frameworks often drive interesting phenomena in research areas such as string theory, quantum gravity, the quantum hall effect, and more recently quantum cosmology. The applications span a myriad of counter-intuitive features, from non-commutativity of geometry to multi-valued dispersion relations. Yet they share a handful of common properties that seem to be at the foundation of non-locality, such as fuzzy size for excitations that increases with momentum~\cite{Seiberg:2000ms}, faster than light signal propagation~\cite{Hashimoto:2000ys}, and a mechanism akin to the Landau levels problem~\cite{Seiberg:1999vs,Bigatti:1999iz}.
In this work, we focus on a recent effort known as Puff field theory (PFT)~\cite{Ganor:2006ub,Ganor:2007qh} that realizes non-local dynamics within a computationally accessible and unitary setting of string theory.

\subsection{The setup}

The particular details of the consistent truncation of the full string theory that leads to the PFT sets the theory apart from other similar attempts such as Non-Commutative Yang-Mills (NCSYM), Non-Commutative Open Strings (NCOS), and Dipole theories~\cite{Seiberg:2000ms,Seiberg:1999vs,Maldacena:1999mh,Hashimoto:1999ut,Hashimoto:1999yj,Klebanov:2000pp,Bergman:2000cw}. Regardless, the general strategy in all these cases is the same: one takes inherently non-local degrees of freedom in string theory such as D-branes, turns on a background flux to polarize the partons or twists the geometry, and identifies a low energy limit that decouples the theory from gravity yet leaving it with the non-local character of the parent string theory. In the case of the PFT, one focuses on D3 branes in a Melvin universe with an RR 5-form background flux. The resulting decoupled theory exhibits non-locality with varied amounts of supersymmetry and R-symmetry. For the PFT of interest in this paper, we deal with a $3+1$ dimensional PFT with $\mathcal{N}=2$ supersymmetry and $ U(1)\times U(2)$ R-symmetry.

A hallmark of the phenomenon of non-locality in string theory is a mixing of high and low energy dynamics. For energies low enough compared to the scale of non-locality, the size of an excitation in the theory is characterized by its Compton wavelength; however, at higher energies one finds that larger energies correspond to bigger physical sizes for the excitations.
For most NCSYM, NCOS, or Dipole theories, this fuzzing of the excitations occurs in a subset of the physical dimensions of the theory and breaks Lorentz symmetry in a manner that is phenomenologically undesirable. In our PFT, the excitations expand in all three space directions, breaking Lorentz symmetry only to $SO(3)$.

Unfortunately, a Lagrangian formulation of the PFT is still missing. However, a strong coupling computational framework exists through gravitational holography~\cite{Maldacena:1997re,Witten:1998qj}. Hence, we focus on the strongly coupled regime of the PFT and attempt to unravel the non-local aspects of its dynamics. The IIB gravitational dual of the PFT may be split into two five dimensional parts $\mathcal{M}_5$ and $\mathcal{C}_5$; with the PFT living in a four dimensional subspace of $\mathcal{M}_5$, and $\mathcal{C}_5$ being a five dimensional compact manifold with varying size as one moves in the  $\mathcal{M}_5$. We denote the coupling in the theory by $G\gg 1$, with $\Delta$ being the length scale of non-locality. Non-local aspects of the theory are expected to arise for operators that carry R-charge. Hence, we employ probes in the dual bulk spacetime with the associated momenta. These consist of excitations arising from Kaluza-Klein modes on the $\mathcal{C}_5$. We find that we can approximate the analysis through the use of geodesics, while still having back-reaction effects and the optical limit under control.

And through the use of null and spacelike geodesics, we explore UV-IR mixing and equal-time correlators in the PFT. Our analysis leads to a sequence of novel and surprising observations that set the PFT apart from other non-local theories and renders it even more compelling. The results may be summarized as follows:

\begin{itemize}
	\item We find that the bulk space has an apparent horizon that divides it into two regions and where the rate of expansion of null geodesics vanishes. Bousso's criterion~\cite{Bousso:1999xy} for holography suggests that the two regions are holographically encoded onto the same holographic screen. Other works~\cite{Ryu:2006bv,Hubeny:2007xt} suggest that such screens can also play the role of entangling the quantum theories on either side. In this setting, we find that one has a quantum entanglement of UV and IR dynamics.  
	\item We propose a new holographic prescription for mapping observables in this bulk space to observables in the PFT. This involves separating the notion of integrating out high energy modes from that of integrating out high momentum modes. Thus we are left with two different realizations of renormalization group flow, with the UV completion of the PFT sitting at the apparent horizon.
	\item We compute equal-time correlators for various observables, reproducing the expected low energy or low momentum results while adding new physics at high energies or high momenta. The correlators smoothly interpolate between power law and exponential behavior as one moves from low to high energies.
	\item We find that the PFT exhibits non-locality at an energy scale $G^{-1/3}\Delta^{-1}$ at strong coupling. Hence, the effects of non-locality are enhanced.
	\item We develop a new relation between momentum and spatial support size in the PFT that associates a single momentum scale with two sizes, and vice versa.
	\item We compute a c-function for the PFT using null geodesics in the bulk. We find that the PFT behaves very differently from NCSYM and NCOS theories, exhibiting a divergence in the c-function at the apparent horizon and violating the monotonicity condition in the UV. We confirm that the latter arises from a violation of the null energy condition in the bulk in certain regions.
	\item Bringing together all these observations, we apply these results to a toy cosmological scenario. We assume that at some early time in the history of the universe, the primordial plasma was described by a theory similar to strongly coupled PFT. Hence, equal-time correlators in the PFT determine the initial conditions of fluctuations of the plasma, which then evolves through inflation and leads to structure formation. We compute the power spectrum that results from such a scenario and find that signature of stringy physics can indeed arise at small length scales in the Cosmic Microwave Background (CMB). This is due to a combination of an enhancement of non-local effects at strong coupling and the stretching of small scale structure though the violent expansion of space.
\end{itemize}

The outline of the paper goes as follows. We start in Section 2 with setting up the gravitational background that is holographically dual to strongly coupled PFT. We determine the various conditions on the parameters that assure that the computation is reliable. In Section 3, we first look at the expansion/contraction rate of null geodesics in the bulk identifying an apparent horizon and entanglement screen. We then proceed to developing the UV-IR relations that relate energy/momentum scale in the PFT to extent in the bulk. In the last subsection, we present the prescription that allows the computation of correlators in the PFT using geodesics of the bulk space. We also present a mechanism for entangling `high and low energy operators' so as to naturally reproduce these observables. In Section 4, we present analytical asymptotics and numerical plots of various equal-time correlators. Section 5 discusses a cosmology application of the results. This includes embedding the setup into the so-called minimal approach of structure formation; one then uses the correlators from the PFT to compute the power spectrum of scalar fluctuations. The Appendices describe a computation of the c-functions for the PFT, NCSYM, and NCOS theories, contrasting the results, as well as the development of a non-local realization of Fourier transforms, relating momentum to length scale in the PFT.

\section{Strong coupling regime of Puff field theory}

PFT may be thought of as the $3+1$ dimensional worldvolume theory for $N$ D3 branes in a Melvin universe supported by a background 5-form RR flux~\cite{Ganor:2006ub,Ganor:2007qh}. The theory is parameterized by the number of D3 branes $N$, a coupling constant $G$, and a scale of non-locality $\Delta$ which has dimensions of length. In~\cite{Ganor:2007qh}, the theory is also compactified on a 3-torus. One can realize different versions of PFT with a varying number of supersymmetries. For our purposes, we will only consider the decompactified theory with $\mathcal{N}=2$ supersymmetry in $3+1$ dimensions. At low energies, this PFT is expected to flow to $\mathcal{N}=4$ Super Yang-Mills (SYM). 

The PFT of interest has full $SO(3)$ symmetry while being a non-local theory. The R-symmetry is broken down from $SO(6)$ to $U(1)\times U(2)$. Excitations with large momenta and with R-charge are expected to fuzz out in all three space directions to a volume proportional to $\Delta^3$. While a full Lagrangian description of PFT is still missing, we will only be concerned with the strong coupling regime of the theory; the latter is accessible through holography by studying supergravity excitations in a particular curved background. This background was obtained in~\cite{Ganor:2007qh}. The IIB string metric is given by 
\begin{eqnarray}
	\label{eq:thebigmetric} ds_{str}^2&=& \alpha' K^{1/2} \left[ -H^{-1} dt^2 +dV^2+ V^2 d\psi^2 +\frac{V^2 \cos^2\psi}{4} \left( d\theta^2+\sin^2 \theta\, d\varphi^2\right) +V^2 \sin^2 \psi\, d\chi^2 \right] \nonumber \\
	&+& \alpha' K^{-1/2} \left[ \sum_i dx_i^2 + H\, V^2 \cos^2\psi\, \left( d\phi -\frac{1}{2} \left( 1-\cos\theta \right)d\varphi + \Delta^3 H^{-1} dt \right)^2 \right] \ .
\end{eqnarray}
The worldvolume of the D3 branes spans the coordinates $x_i$ with $i=1,2,3$ and the time direction; the holographic direction associated with energy scale in the dual PFT is denoted by $V$; and the remaining five directions are compact and denoted by $\{\phi$, $\varphi$, $\psi$, $\theta, \chi\}$. The angles $\theta$, $\phi$, and $\varphi$ parameterize an $S^3$ which may be viewed as a Hopf fibration with $\phi$ labeling the fiber direction and the base $S^2$ described by $\{\theta,\varphi\}$. The two functions $H(V)$ and $K(V)$ appearing in the metric are given by
\begin{equation}
	H=\frac{4\pi g_s N}{V^4}\ \ \ \mbox{and}\ \ \ K=\frac{4\pi g_s N}{V^4}+\Delta^6 V^2 \cos^2\psi \ ,
\end{equation}
where $g_s$ is the IIB string coupling, $N$ is the number of D3 branes, and $\Delta$ is the scale of non-locality. The dilaton profile is constant 
\begin{equation}
	e^\Phi = g_s 
\end{equation}
but there is a non-trivial 5-form RR flux that supports the geometry. The latter is of no relevance to our discussion as we will concentrate on supergravity probes with zero RR charge.

Furthermore, we focus on the five dimensional subspace spanned by $t$, $x_i$, and $V$ by considering processes involving no angular dynamics 
\begin{equation}
	\mbox{All angles }\psi, \theta, \varphi, \chi, \phi=\mbox{Constants} 
\end{equation}
with $\psi=0$ for convenience (note that this point is a coordinate singularity of~(\ref{eq:thebigmetric})). Timelike or null geodesics confined to this cross-section of the space carry momentum in the $\phi$ and $\varphi$ directions because of the twist in the geometry, as seen through the $t-\phi$ and $t-\varphi$ cross-terms of the metric. These are seen by the PFT as probes with R-charge. For example, taking $\theta=0$ for simplicity, $\partial_\phi$ is a Killing vector and geodesics stationary in the angle directions still carry momentum in the $\phi$ direction given by
\begin{equation}\label{eq:phimomentum}
	\mbox{Momentum in $\phi$ direction} \propto g_{t\phi} \frac{dt}{d\tau}
\end{equation}
which scales as $\Delta^3$ for small $\Delta$.
More importantly, Kaluza-Klein modes in the $\phi$ direction generate supergravity modes of mass
\begin{equation}
	m_{KK}\sim \frac{K^{1/4}}{\sqrt{\alpha'} H^{1/2} V}\ .
\end{equation}
We shall see that these masses are parametrically small in string units. Probes associated with these excitations would then carry R-charge as well.

\begin{figure}
	\begin{center}
		\includegraphics[width=6in]{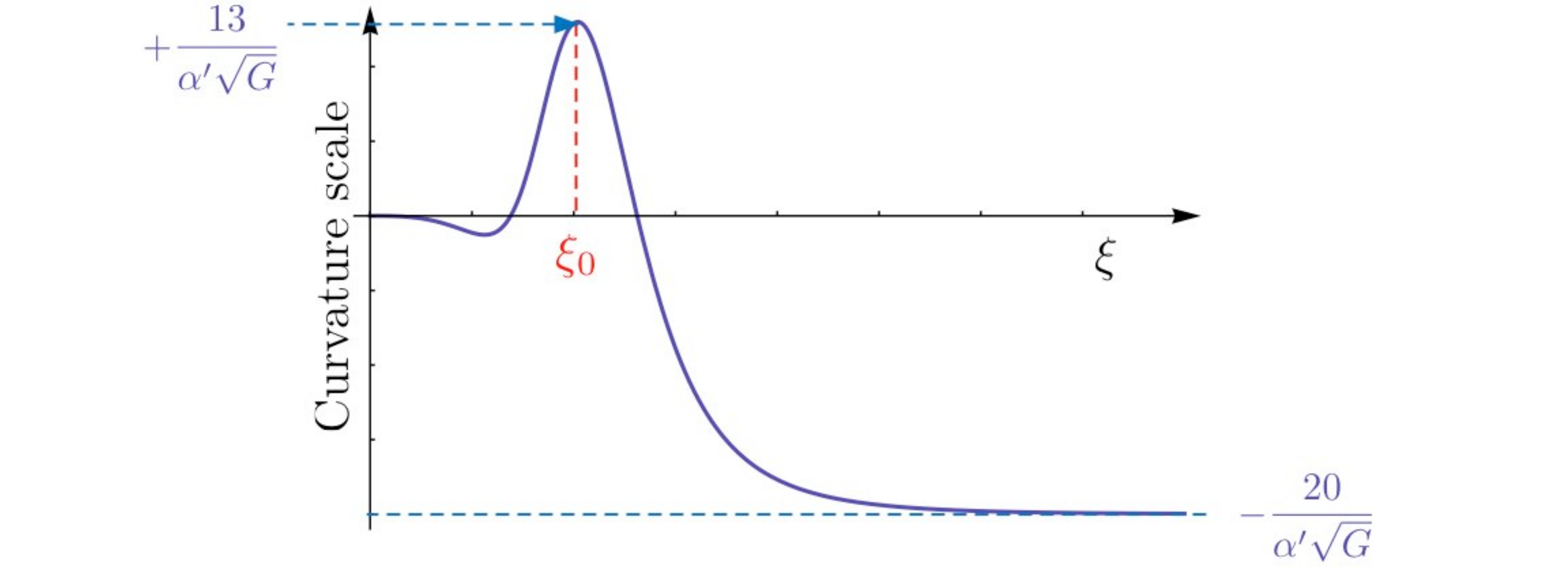}
	\end{center}
	\caption{The Ricci scalar for the bulk spacetime dual to the PFT as a function of the holographic coordinate $\xi$. The throat referred to in the text appears at $\xi=\xi_0=2^{-1/6}$.} \label{fig:curvatureplot} 
\end{figure}

Consider a coordinate transformation 
\begin{equation}
	z=\frac{1}{V} \ ,
\end{equation}
and a rescaling of all five coordinates 
\begin{equation}\label{eq:rescalings}
	\xi\equiv \frac{G^{1/6}}{\Delta} z\ \ \ ,\ \ \ X_i\equiv \frac{x_i}{G^{1/3} \Delta}\ \ \ ,\ \ \ T \equiv \frac{t}{G^{1/3} \Delta} \ ,
\end{equation}
where 
\begin{equation}
	G\equiv 4\pi g_s N \ .
\end{equation}
We will later need to take $N\gg 1$, with $G$ being the effective coupling in the PFT. For the holographic setting to be useful, we then need $G\gg 1$ in addition. In the new coordinates, the metric becomes 
\begin{equation}
	\label{eq:puffmetric} ds_{str}^2=\alpha'\, G^{1/2}\, \left[ -\frac{\xi}{\sqrt{1+\xi^6}} dT^2 +\frac{\sqrt{1+\xi^6}}{\xi^5} d\xi^2 +\frac{\xi}{\sqrt{1+\xi^6}} \sum_idX_i^2\right] \ .
\end{equation}
Note that this metric includes the $dt^2$ term from the twisted part of the metric.
The holographic direction is $\xi$, and there is a throat in the geometry at 
\begin{equation}
	\xi^6=\xi^6_0=\frac{1}{2} 
\end{equation}
as shown in Figure~\ref{fig:curvatureplot}.
The Einstein frame metric has the overall factor $\sqrt{G}$ replaced by $\sqrt{2\pi N}$:
\begin{equation}
	ds_{Ein}^2=g_s^{-1/2} ds_{str}^2 = \frac{\sqrt{4\pi N}}{G^{1/2}} ds_{str}^2 \ .
\end{equation}
Note also that the decoupling limit involves taking $\alpha'\rightarrow 0$ while $G$, $\xi$, $T$, $X_i$ are held fixed.

We want to probe this geometry with geodesics, using them to compute correlation functions in the PFT at strong coupling $G\gg 1$. Hence, we need to assure that the supergravity description and the optical approximation are valid. For small curvatures compared to the string scale, this leads to 
\begin{equation}
	\label{eq:smallcurvatures} G \gg 1\ \ \ \ \ \ \mbox{Small curvatures.} 
\end{equation}
We also need to make sure that the circle along $\phi$ is not too small compared to the string scale 
\begin{equation}
	\label{eq:tduality} \xi \gg G^{-1/6}\ \ \ \ \ \ \mbox{Large circle along $\phi$.} 
\end{equation}
Otherwise, we would need to consider the T-dual geometry. Note that this condition also renders Kaluza-Klein modes associated with this circle light. We also require weak string coupling 
\begin{equation}
	\label{eq:weakstringcoupling} g_s= \frac{G}{4\pi N}\ll 1\ \ \ \ \ \ \mbox{Weak string coupling everywhere.} 
\end{equation}
Using~(\ref{eq:smallcurvatures}), this also implies that one needs $N\gg 1$.
Finally, the optical approximation implied by a geodesic treatment is valid when the wavelength of the bulk probe is much smaller than the local curvature of the space it is moving through 
\begin{equation}
	\label{eq:optical} \frac{1}{\mbox{Momentum scale}} \ll \mbox{Local curvature length scale,} 
\end{equation}
where the momentum of a geodesic probe is to be measured locally as well. The local curvature can be measured through the Ricci scalar 
\begin{equation}
	R=\frac{-20 \xi ^{15}+104 \xi ^9-11 \xi ^3}{\alpha' \, \sqrt{G} \left(\xi ^6+1\right)^{5/2}}\equiv \frac{1}{l_{crv}^2} \ .
\end{equation}
Measured in terms of the coordinate variable $\Delta X$, this length scale is 
\begin{equation}
	\label{eq:curvaturescale} \Delta X_{crv} \sqrt{g_{XX}}=l_{crv}\Rightarrow \Delta X_{crv}=\frac{(1+\xi^6)^{3/2}}{\xi^2\sqrt{-11+104 \xi^{6}-20 \xi^{12}}} \ .
\end{equation}
The momentum canonically conjugate to the $X$ coordinate for a spacelike geodesic probe of interest is given by $P\sim m_0\, g_{XX}\, dX/d\lambda$. We will see later (through equations~(\ref{eq:puffmomcons}), (\ref{eq:prescaling}) and~(\ref{eq:pmax})) that this momentum scale is bounded as in 
\begin{equation}
	0< P\lesssim \sqrt{\alpha'}\,m_0\,G^{1/4} 
\end{equation}
where $m_0\equiv M_0/\sqrt{\alpha'}$ is the rest mass of the probe. The curvature length scale $\Delta X_{crv}$ given by~(\ref{eq:curvaturescale}) is bounded from below (with a single minimum at $\xi\simeq 1.02$ where $\Delta X_{crv}\simeq 0.33$). Equation~(\ref{eq:optical}) then leads to 
\begin{equation}
	\label{eq:opticalapprox} \Delta X_{crv} \gg P^{-1}\Rightarrow M_0 {G}^{1/4}\mathcal{P} \gg 1\ \ \ \ \ \ \mbox{Required by optical approximation}
\end{equation}
where $\mathcal{P}$ is a small numerical scale tuning the momentum in units of the maximum momentum
\begin{equation}
	P\equiv \mathcal{P} M_0\,G^{1/4}\ .
\end{equation}
The important point is in the presence of the $\sqrt{G}$ factor. We see that one can extend the geodesic treatment to probes of mass scale $M_0\ll 1$ and $\mathcal{P}$ parametrically small as long as we have $G\gg 1$. Basically, since the curvature is bounded, we can always push curvature length scales {\em for all} $\xi$ to large enough values to extend the optical approximation to a larger parameter space in $M_0$ and $\mathcal{P}$. This is needed because we want to access small momenta, and we also need to keep back-reaction effects under control
\begin{equation}\label{eq:smallmass}
	M_0\ll 1\ \ \ \ \ \ \mbox{Small back-reaction from probe.}
\end{equation}
For example, one can achieve this by considering a low lying Kaluza-Klein mode of the $\phi$ circle, which is large in string units as required by the T-duality condition~(\ref{eq:tduality}). This also implies that such a probe will necessarily carry R-charge under the $U(1)\times U(2)$ R-symmetry of the PFT.

For the rest of the paper, we focus on the regime delineated by~(\ref{eq:smallcurvatures}), (\ref{eq:tduality}), (\ref{eq:weakstringcoupling}), (\ref{eq:opticalapprox}), and~(\ref{eq:smallmass}). Together, these require very large $N$ and $G$, and avoiding the use of geodesics of very small momenta that probe regions of small $\xi$.

\section{The holographic dictionary for the PFT}

\subsection{Null geodesics and covariant holography}

\begin{figure}
	\begin{center}
		\includegraphics[width=6in]{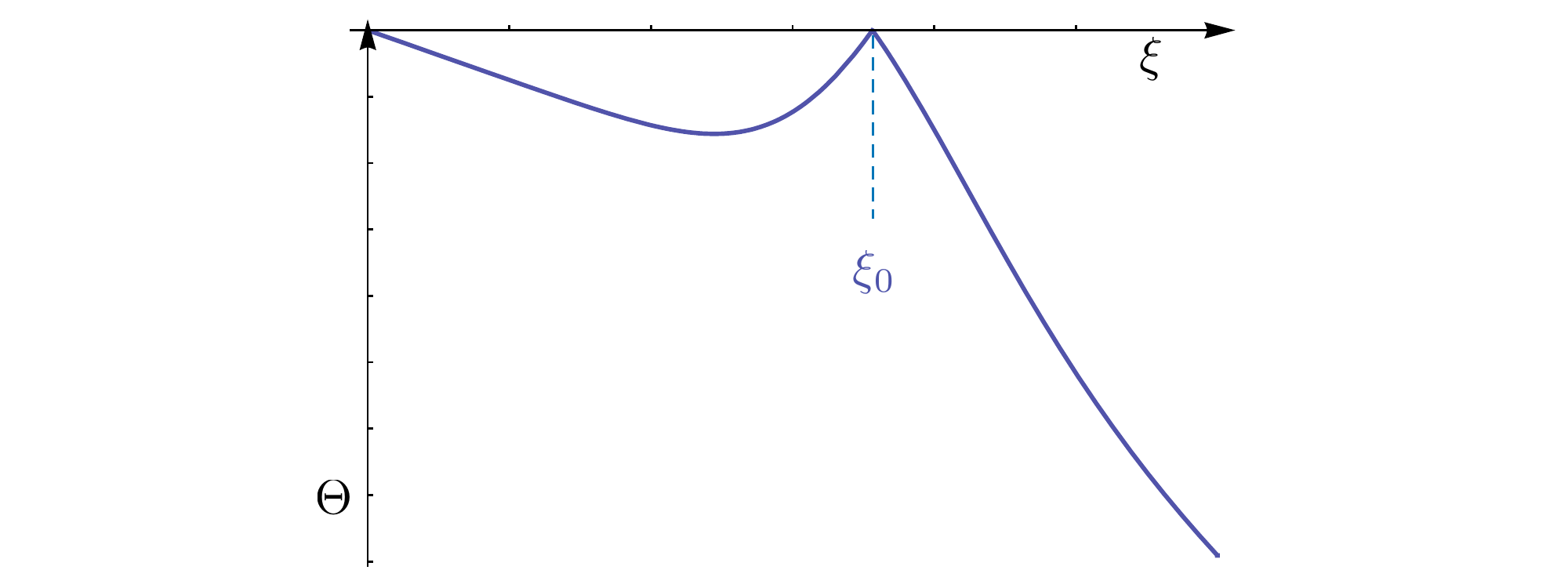}
	\end{center}
	\caption{Convergence rate $\Theta$ of future-directed null geodesics projected away from $\xi=\xi_{0}$ surface.} \label{fig:pufftheta} 
\end{figure}

We start by focusing on a congruence of null geodesics spanning the $T$-$\xi$ subspace of the geometry. $
\partial_T$ being a Killing vector, we have the conservation law 
\begin{equation}
	\frac{dT}{d\lambda}=\frac{E}{\alpha' \sqrt{4\pi N}}\frac{\sqrt{1+\xi^6}}{\xi} 
\end{equation}
where $E$ is a constant. We then write the tangent to future directed null geodesics in the $T$-$\xi$ plane as 
\begin{equation}
	t^a=\frac{1}{\sqrt{4\,\pi\, N}\alpha'}\left( \frac{\sqrt{1+\xi^6}}{\xi} 
	\partial_t \pm \xi^2 
	\partial_\xi \right)\ .
\end{equation}
The second fundamental form is 
\begin{equation}
	K_{ab}=\nabla_{(a}t_{b)} 
\end{equation}
whose trace $\Theta$ measures the rate of expansion/contraction of the geodesics 
\begin{equation}
	\label{eq:thetheta} {\Theta} = {K_a^a}= \pm \frac{1}{\alpha'}\frac{3}{2\sqrt{4\,\pi\, N}}\, \frac{\xi}{1+\xi^6}\left( 1-\frac{\xi^6}{\xi_{0}^6}\right) \ ,
\end{equation}
where the upper plus sign corresponds to future directed geodesics moving towards increasing $\xi$ while the lower minus sign corresponds to future directed geodesics towards decreasing $\xi$. We then conclude that 
\begin{equation}
	\begin{array}{ll}
		\Theta < 0 & \mbox{For future directed geodesics projected away from the $\xi=\xi_0$ point.} \\
		\Theta =0 & \mbox{At $\xi=\xi_0$.} 
	\end{array}
\end{equation}
The asymptotics of~(\ref{eq:thetheta}) are 
\begin{equation}
	\label{eq:thethetaasympt} \left|{\Theta}\right|\simeq \left\{ 
	\begin{array}{ll}
		\xi / 2 & \xi\ll\xi_{0} \\
		\xi & \xi\gg\xi_{0} 
	\end{array}
	\right. 
\end{equation}
and Figure~\ref{fig:pufftheta} shows a plot.
The $\xi=\xi_0$ surface is then an {\em apparent horizon} of the geometry. Note that when $\Delta\rightarrow 0$, with $\xi_0= G^{1/6} z_0/\Delta$, this feature of the geometry disappears as $z_0\sim \Delta \rightarrow 0$. 

\begin{figure}
	\begin{center}
		\includegraphics[width=6in]{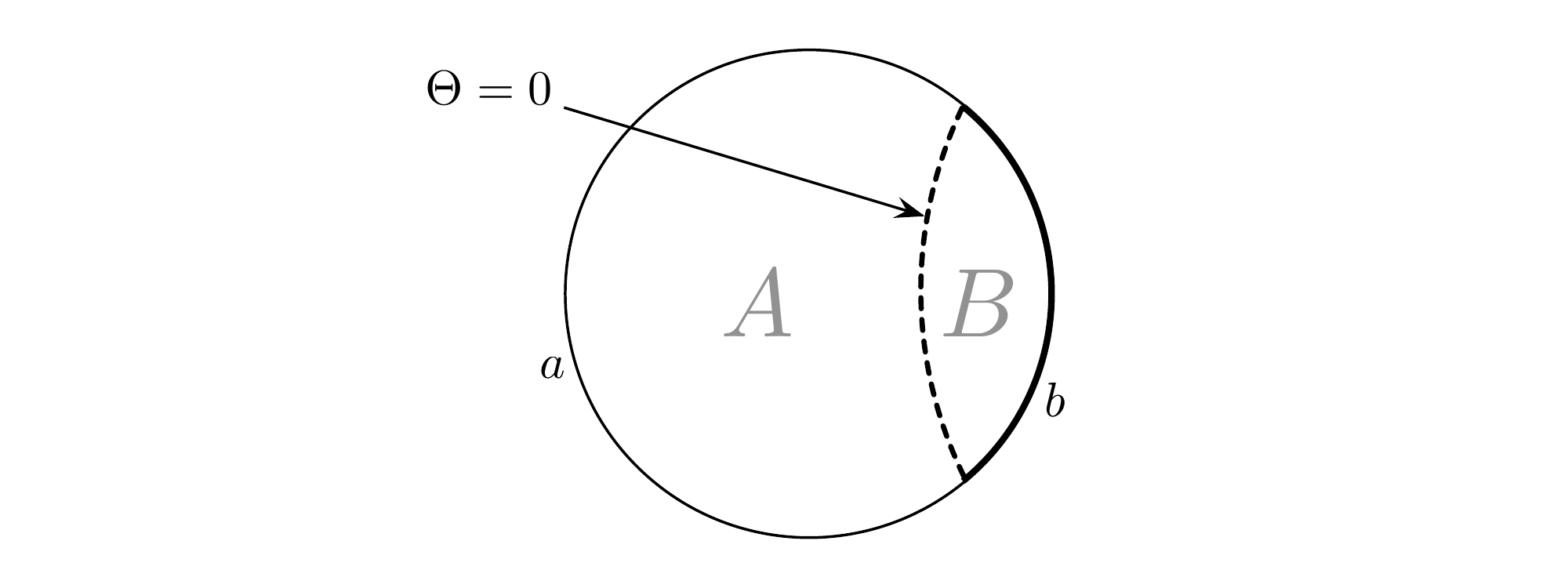}
	\end{center}
	\caption{A cartoon of the covariant entanglement proposal of~\cite{Ryu:2006bv,Hubeny:2007xt}. The $\Theta=0$ entanglement screen separates the bulk into two regions A and B, and the boundary into two regions $a$ and $b$. It is proposed that the entanglement entropy of region a or b is given by the area of entanglement screen.} \label{fig:entanglement} 
\end{figure}

Bousso's criterion~\cite{Bousso:1999xy} for holography would then suggest that there exists a holographic screen at $\xi_0$ encoding entropy and information from two regions on either side of the screen. This constitutes a novel realm of holography in string theory. A related setting  was recently explored in~\cite{Ryu:2006bv,Hubeny:2007xt} (from other work on holographic entanglement, see for example~\cite{Hawking:2000da,Maldacena:2001kr,Brustein:2005vx,Buniy:2005au}). The authors of these works considered the computation of entanglement entropy of a conformal field theory by tracing over states lying outside a region interest. In the holographic dual of this process, they suggested to slice the bulk with a screen where the congruence of null geodesics vanishes - with the boundaries of this screen matching onto the boundaries of the region of interest in the dual theory (see Figure~\ref{fig:entanglement}).
The suggestion is that the surface where $\Theta$ vanishes divides two bulk regions that map onto a pair of entangled density of states. 
For the PFT, the presence of an apparent horizon is suggestive that an entanglement mechanism may be at work as well. In the Appendices, we demonstrate that this phenomenon is absent from NCSYM and NCOS theories. In the next subsection, we explore UV-IR relations to help us unravel this intricate phenomenon apparently particular to the PFT.

\subsection{UV-IR relations}  \label{sec:uvir}

To determine the relation between energy scale in the PFT and the holographic coordinate $\xi$, we could look at the finite temperature version of our geometry. One then easily finds that temperature in the PFT scales as~\cite{Ganor:2007qh}
\begin{equation}
	\label{eq:temperature} \mbox{Temperature}\simeq \frac{1}{\Delta\, G^{1/3}}\frac{1}{\xi_H}\equiv \omega_t\ ,
\end{equation}
where $\xi=\xi_{H}$ is the location of the horizon in the corresponding geometry. The surface gravity and hence the temperature are not affected by the cross terms $d\phi-dt$ and $d\varphi-dt$ of the metric, hence the simple form of~(\ref{eq:temperature}). 
We will henceforth refer to the $\xi\ll\xi_0$ region as the UV domain, and the $\xi\gg\xi_0$ as the IR domain.

This simple UV-IR relation however masks the rich non-local substructure of the theory.
To better understand the problem, consider spacelike geodesics in the $\xi-X_i$ subspace.
Using the $SO(3)$ symmetry, we can always align such a geodesic within a chosen two dimensional plane, which we parameterize for simplicity by $\{\xi,X\}$. The Killing vector $\partial_X$ leads to the conservation statement 
\begin{equation}
	\label{eq:puffmomcons} \frac{dX}{d\lambda}=\frac{p}{\sqrt{\alpha'} G^{1/2}}\frac{\sqrt{1+\xi^6}}{\xi}\ ,
\end{equation}
where $p$ is a constant related to momentum. And the geodesic being spacelike, we have 
\begin{equation}
	\label{eq:puffspacenorm} 1=\alpha'\, G^{1/2}\, \left[ \frac{\sqrt{1+\xi^6}}{\xi^5} \frac{d\xi^2}{d\lambda^2} +\frac{\xi}{\sqrt{1+\xi^6}} \frac{dX^2}{d\lambda^2}\right]. 
\end{equation}

\begin{figure}
	\begin{center}
		\includegraphics[width=6in]{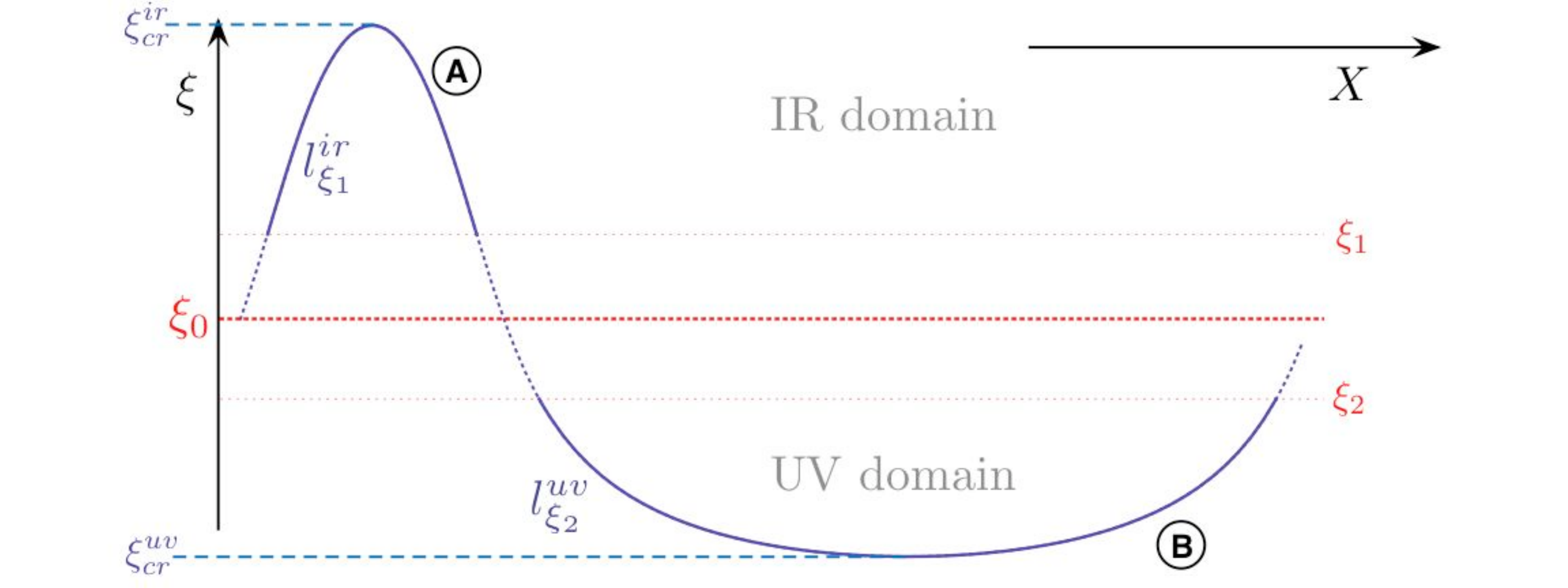}
	\end{center}
	\caption{Oscillating spacelike geodesics with $P=0.7$ in the $\xi-X$ plane. The length of the segment labeled (A) is denoted in the text by $l_{\xi_1}^{ir}(P)$; the length of the segment (B) is denoted by $l_{\xi_2}^{uv}(P)$.} \label{fig:geodesicplot} 
\end{figure}

\begin{figure}
	\begin{center}
		\includegraphics[width=6in]{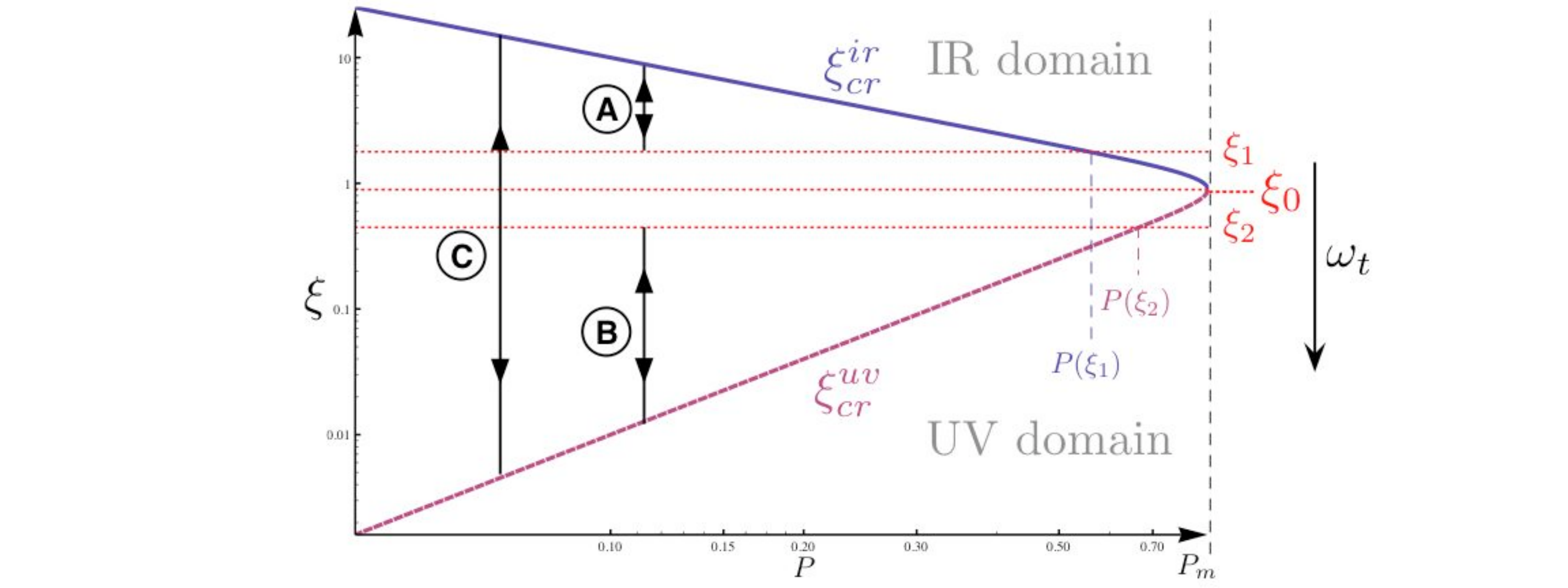}
	\end{center}
	\caption{Log-log plot of equation~(\ref{eq:puffmomxi}) relating the turn-around critical point $\xi_{cr}$ to momentum $P$.} \label{fig:critpoints} 
\end{figure}

Combining (\ref{eq:puffmomcons}) and (\ref{eq:puffspacenorm}), we solve for the slope of geodesics in the $\xi-X$ plane 
\begin{equation}
	\label{eq:puffgeoslope} \frac{dX}{d\xi}=\pm\frac{P (1+\xi^6)^{3/4}}{\xi^3 (\xi-P^2 \sqrt{1+\xi^6})^{1/2}} \ ,
\end{equation}
with the new momentum parameter rescaled as 
\begin{equation}\label{eq:prescaling}
	P \equiv \frac{p}{{G^{1/4}}} \ .
\end{equation}
We then have turn-around points where $d\xi/dX = 0$. Using (\ref{eq:puffgeoslope}), one finds 
\begin{equation}
	\label{eq:puffmomxi} \frac{d\xi}{dX}=0\Rightarrow P(\xi_{cr})=\frac{\xi_{cr}^{1/2}}{(1+\xi_{cr}^6)^{1/4}} \ .
\end{equation}
We can also invert (\ref{eq:puffmomxi}) to find the turn-around point at a given momentum 
\begin{equation}
	{\left(\xi^{uv,ir}_{cr}\right)}^2 = \frac{{P_{m}}^2}{2^{1/3}P^2} \left[\cos\frac{\chi}{3}\pm \sqrt{3}\sin \frac{\chi}{3}\right]\ \ \ \mbox{with}\ \ \
	\tan \chi \equiv \frac{\sqrt{1-P^{12}/P_m^{12}}}{P^6/P_m^6} \ ,
\end{equation}
where the IR label refers to the top choice of sign, and the UV label to the bottom choice.
Geodesics then oscillate between the two critical points for fixed momentum $P$ as shown in Figure~\ref{fig:geodesicplot}.
Equation~(\ref{eq:puffmomxi}) implies a maximum value for the momentum $P$, which arises in the limit of $\xi \rightarrow \xi_0$ 
\begin{equation}\label{eq:pmax}
	P_{m}=\frac{2^{1/6}}{3^{1/4}} 
\end{equation}
and 
\begin{equation}
	0<P<P_{m} 
\end{equation}
as can be seen in Figure~\ref{fig:critpoints}.

Consider the PFT with a UV cutoff $\xi_1>\xi_0$ as shown in Figures~\ref{fig:geodesicplot} and~\ref{fig:critpoints}. If we take $\xi=\xi_1$ as a holographic screen, we have $\Theta<0$ for a congruence of null geodesics {\em projected from $\xi_1$} only for $\xi>\xi_1$. Hence, it is the upper segment that is a candidate for holographic encoding -  for example through the representative geodesic labeled (A) in the figures. The maximum momentum that may be probed at this UV cutoff is given by $P<P(\xi_1)$.
In this regime, as we increase the UV cutoff $\omega_t$ and hence decrease $\xi_1$, we may probe the PFT at large momenta $P(\xi_1)$. This synchs well with our intuition from local field theories. 

However, if the UV cutoff is chosen in the UV region at say $\xi=\xi_2$ (as shown in Figure~\ref{fig:critpoints}), we now have $\Theta<0$ for null geodesics projected from $\xi_2$ for the {\em lower} segment $\xi<\xi_2$ instead; and we still have a maximal momentum $P<P(\xi_2)$. The region available for holographic encoding is given by $\xi<\xi_2$ and may be probed by the representative geodesic labeled (B) in the figure. However, higher UV cutoffs $\omega_t$ correspond to smaller momenta $P(\xi_2)$! This can only make sense in the renormalization group spirit if we divorce the concept of integrating out high energy modes from that of integrating out high momentum modes.

The discussion leads us to the following modified picture for renormalization group in this non-local theory. If one wants to probe the PFT up to a certain momentum $P_0$, this picks two corresponding values in the bulk for holographic screens, say at $\xi_{cr}^{uv}(P_0)$ and $\xi_{cr}^{ir}(P_0)$. We are then to integrate out physics in the bulk lying in the region $\xi_{cr}^{uv}(P_0)<\xi<\xi_{cr}^{ir}(P_0)$.

On the other hand, we may also be interested in probing the theory after integrating out high energy modes instead of large momentum scales. In the PFT, these two are not equivalent as we argued above. We may still construct probes in this more conventional sense by focusing only on geodesics of the type labeled (A) in~Figure~\ref{fig:critpoints}, as long as we keep the energy UV cutoff bounded $\Lambda<1/(G^{1/3} \Delta)$, {\em i.e.} $\xi_1>\xi_0$. We will elaborate on these possibilities through explicit computations in the next section. 

Note that these two prescriptions necessarily involve integrating out behind the apparent horizon. Such a process may lead to entangled observables - as may be expected from~\cite{Ryu:2006bv,Hubeny:2007xt}. 
It is now also clear that one should think of the UV completion of the theory as sitting at the throat $\xi=\xi_0$. 

If we write the Compton wavelength of an excitation in the PFT of momentum $P$ as $1/P$, we see that this length scale is bounded from below: at maximal momentum $P_m$ and $\xi=\xi_0$, one probes this minimal length scale which scales as
\begin{equation}
	\Delta X\sim \frac{1}{P_m}\Rightarrow\Delta x\sim G^{1/3} \Delta \ .
\end{equation}
Note that this scale is bigger than $\Delta$ at strong coupling. The interactions in the theory seem to enhance the effects of non-locality. 
We can use~(\ref{eq:puffmetric}) to easily see that that size $\Delta X_\xi$ of a probe or supergravity fluctuations goes as 
\begin{equation}
	\label{eq:deltaL} {\Delta X_\xi} \simeq\frac{1}{\sqrt{g_{XX}}|_{\xi}} \sim \frac{(1+\xi^6)^{1/4}}{\xi^{1/2}}\ .
\end{equation}
This is simply the redshift effect and leads to $\Delta X_{\xi_{cr}}\sim 1/P(\xi_{cr})$ as expected. Given that we use the momentum label $P$ for {\em spacelike} geodesics, it is more suitable to think of this parameter as a measure of spatial resolution. However the association of fuzz size to momentum in the PFT is somewhat trickier than this simplest viewpoint. In the Appendix, we elaborate on this issue exploring a non-local version of the Fourier transform relating size in $X$ to $P$. In the main text, we confine the discussion to momentum space for simplicity. 

\begin{figure}
	\begin{center}
		\includegraphics[width=7in]{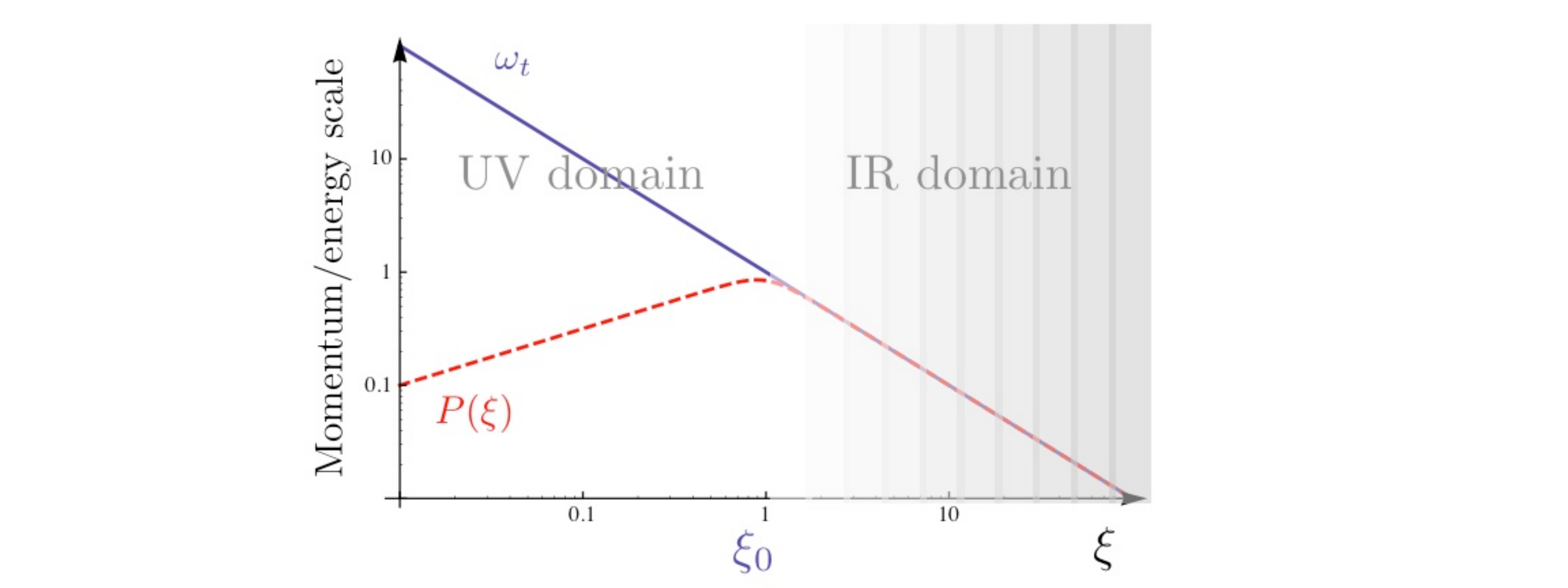}
	\end{center}
	\caption{Log-log plots of the various UV/IR relations described in the text.} \label{fig:uvirenergy} 
\end{figure}

In summary, Figure~\ref{fig:uvirenergy} shows plots of the two UV/IR relations encountered in the text.

\subsection{Correlators}

In general, one expects that the correlators of the PFT at strong coupling are to be related to the length of geodesics in the dual bulk spacetime. In the current setting, this map is however more elaborate. Consider specifically spacelike geodesics of the geometry, with both endpoints fixed at the same $\xi = \xi_{1,2}$ at equal time (see Figure~\ref{fig:critpoints}). We want to relate the length of such a geodesic to the equal time correlation function of an operator in the PFT - along with a UV cutoff related to $\xi_{1,2}$. 
We are then interested in the length for a round-trip from $\xi_{1,2}$ to $\xi_{cr}$ and back to $\xi_{1,2}$, given by 
\begin{eqnarray}
	\label{eq:lengthintegral} l_{\xi_{1,2}}&=&2\,\sqrt{\alpha'} G^{1/4} \left|\int_{\xi_{1,2}}^{\xi_{cr}} d\xi \left[ \frac{\sqrt{1+\xi^6}}{\xi^5} +\frac{\xi}{\sqrt{1+\xi^6}} \frac{dX^2}{d\xi^2}\right]^{1/2} \right| \nonumber \\
	&=&2\,\sqrt{\alpha'} G^{1/4} \left|\int_{\xi_{1,2}}^{\xi_{cr}} d\xi \frac{\left(1+\xi ^6\right)^{1/4}}{\xi^2 \left(\xi -P^2 \sqrt{1+\xi^6}\right)^{1/2}}\right| \ ,
\end{eqnarray}
where the factor of $2$ is added since geodesics must go to the critical point and then back, and we may take $\xi_{1,2}\rightarrow \xi_0$ if desired. We will henceforth denote the length of the geodesic that bounces off $\xi_{cr}$ in the IR domain as $l^{ir}$, and the one bouncing in the UV domain as $l^{uv}$.

Next, we propose an extension to the holographic dictionary for computing correlators, one that is suited to the setting of the PFT. We do this with the premise that the low energy or small momentum regimes of all observables reproduce the expected results from a local field theory, and that the proposal accounts for our analysis of UV-IR relations and congruences of null geodesics. For a probe of mass $m_0$ associated with a Kaluza-Klein mode, we heuristically have the following natural elements to map onto observables: $e^{-m_0 l_{ir}}$, $e^{-m_0 l_{uv}}$, $e^{-m_0 (l_{ir}+l_{uv})}$ , and $e^{-m_0 l_{ir}}+e^{-m_0 l_{uv}}$. It is useful to keep in mind that - as we shall see in the next section - we always have $l_{ir}<l_{uv}$, with $l_{ir}\ll l_{uv}$ in the low momentum regime. Furthermore, the presence of an entanglement screen between the UV and the IR regimes suggests that we may need a product structure for the Fock space.

Let it be the case that at least a sector of PFT has its Fock space represented by a product of {\em two} Fock spaces - with the vacuum of one denoted by $\left| IR\right>$ while the vacuum of the other being $\left| UV\right>$. We will think of the $\left| IR\right>$ state as dual to the bulk spacetime with $\xi>\xi_0$; while $\left| UV\right>$ is dual to the patch of spacetime bounded by $0<\xi<\xi_0$. A vacuum configuration of the PFT may have the form
\begin{equation}
	\left| 0\right> \equiv \left| IR\right>\times \left| UV\right>\ ;
\end{equation}
or perhaps one should consider instead
\begin{equation}
	\left| \Psi\right> \equiv \alpha \left| IR\right>\times \left| UV\right>+\beta \left| UV\right>\times \left| IR\right>
\end{equation}
where $\alpha$ and $\beta$ are complex constants satisfying $|\alpha|^2+|\beta|^2=1$. We now want to find a natural map of various PFT vevs  onto geodesic length computations in the bulk; and we want to do this in a way to make sense of our previous observations with regards to holography and UV-IR relations.

Consider first the density matrix
\begin{equation}
	\rho_{ir} \equiv\mbox{Tr}_2 \left| 0\right>\left< 0\right|
\end{equation}
where the trace is taken over the second Fock space. This tracing corresponds in the bulk to integrating out dynamics in the region $\xi<\xi_0$, behind the holographic screen. To construct a correlator, we then write
\begin{equation}
	\label{eq:corrir} C_1^\Lambda\equiv \mbox{Tr}_{\Lambda}\left[\mathcal{O}^\dagger(P)\mathcal{O}(P)\rho_{ir}\right]\sim e^{-m_0\, l^{ir}_{\xi_1}(P)} \ ;
\end{equation}
where this involves tracing over part of the $\xi>\xi_0$ region, up to the UV cutoff scale $\Lambda$. This generates the usual renormalization flow. 
In the IR limit $\Lambda\rightarrow 0$, the space becomes $AdS_5\times S^5$ and we can associate with $\mathcal{O}$ the weight~\cite{Witten:1998qj,Balasubramanian:1998sn,Balasubramanian:1998de}
\begin{equation}
	h_+\simeq \frac{M_0 G^{1/4}}{2}
\end{equation}
in view of~(\ref{eq:opticalapprox}) $M_0 G^{1/4}\gg 1$, where $M_0\equiv \sqrt{\alpha'} m_0$. We then expect that the small energy (or small momentum) limit of this correlator would scale as $P^{2 M_0 G^{1/4}}$. Note also that $M_0$ remains finite in the decoupling limit.
Furthermore, one needs to make sure that $\Lambda<\Lambda_m$ where
\begin{equation}
	\Lambda_m\equiv \frac{1}{G^{1/3} \Delta}
\end{equation}
is the energy scale at the throat; otherwise, we cannot make sense of this prescription.

We can also consider the density matrix
\begin{equation}
	\rho_{uv} \equiv \mbox{Tr}_1 \left| 0\right>\left< 0\right|
\end{equation}
where the trace is now taken over the first `IR' Fock space. And we write a corresponding correlator
\begin{equation}
	\label{eq:corruv} C_1^{P_0}\equiv \mbox{Tr}_{P_0}\left[\mathcal{O}^\dagger(P)\mathcal{O}(P)\rho_{uv}\right]\sim e^{-m_0\, l^{uv}_{\xi_2}(P)} \ ;
\end{equation}
Note that the momentum cutoff approach requires us again to restrict the cutoff, now by $P_0<P_m$, the maximum momentum. 

So far, we considered correlators of what we may term pure combinations. We may also ask about mixed states such as $\left| \Psi\right>$ and the resulting entangled vevs. We first specify a momentum cutoff $P_0$; and then trace over part of the bulk space $\xi_1<\xi<\xi_2$ (see Figure~\ref{fig:critpoints}). This involves integrating out behind the holographic screen as well as up to the UV cutoffs $\xi_1$ and $\xi_2$. The natural observable one can construct would start with the density matrix
\begin{equation}
	\rho_{ent} \equiv\mbox{Tr}_2 \left| \Psi\right>\left< \Psi\right|\ ;
\end{equation}
And then we write the correlator
\begin{equation}
	\label{eq:corrmixed} C_2^{P_0}\equiv \mbox{Tr}_{P_0}\left[\mathcal{O}^\dagger(P)\mathcal{O}(P)\rho_{ent}\right]\sim |\alpha|^2 e^{-m_0\, l^{ir}_{\xi_1}(P)}+|\beta|^2 e^{-m_0\, l^{uv}_{\xi_2}(P)}
\end{equation}
where $P(\xi_1)=P(\xi_2)=P_0$. 
We however would not be able to distinguish this case from the one that employs the density matrix 
\begin{equation}
	\rho_{ent} \equiv\mbox{Tr}_1 \left| \Psi\right>\left< \Psi\right|
\end{equation}
instead. This then requires $|\alpha|=|\beta|=1$ for consistency.
Here, we are implicitly mapping expectation values such as $\left< IR\right|\mathcal{O}^\dagger\mathcal{O}\left| UV\right>$ to geodesics that straddle the throat in the geometry; since this region is integrated out, we expect no contribution from such bits of the geodesics.

Finally, we may consider the vev
\begin{equation}
	C_3^{P_0}\equiv \left< 0\right|	\mathcal{O}^\dagger(P)\mathcal{O}(P)\left| 0\right>\ ;
\end{equation}
We would then focus on the two geodesics labeled (A) and (B) in Figure~\ref{fig:critpoints} again. For $P_0=P_m$, we would however get
\begin{equation}
	\label{eq:fullcorrmin} 
	C_3^{P_m}\sim e^{-m_0\, l^{tot}(P)} 
\end{equation}
where $l^{tot}(P)$ is the total length of a geodesic for a round-trip from $\xi_{cr}$ back to the same $\xi_{cr}$ at fixed momentum $P$, as shown by the geodesic labeled (C) in Figure~\ref{fig:critpoints}.

The question remains as to which of the two states $\left| 0\right>$ or $\left| \Psi\right>$ corresponds to the true vacuum of the PFT, dual to the whole of the bulk spacetime. 
In the next Section, we explore numerical and asymptotic forms of these correlation functions and we will see how this prescription can reproduce expected results at low energies and adds new physics at higher energies only. We will hence be lead to guess that the vacuum of the PFT is given by $\left| \Psi\right>$ with $\alpha=\beta=1$.

\section{Numerical results and asymptotics} \label{sec:detailed}

The critical points $\xi_{cr}$ admit straight-forward asymptotic forms by inverting the series expansion of (\ref{eq:puffmomxi}) in the $\xi >> \xi_0$ and $\xi \sim \xi_0$ limits. To leading order, 
\begin{equation} \label{eq:xicritir}
	\xi^{ir}_{cr}= \left\{ 
	\begin{array}{ll}
		P^{-1} & P\ll P_{m} \\
		{\xi_0 \left ( 1 + \sqrt{1 - \frac{P}{P_{m}}} \right )} & P\sim P_{m} 
	\end{array}
	\right. 
\end{equation}
and 
\begin{equation} \label{eq:xicrituv}
	\xi^{uv}_{cr}= \left\{ 
	\begin{array}{ll}
		P^2 & P\ll P_{m} \\
		{\xi_0 \left ( 1 - \sqrt{1 - \frac{P}{P_{m}}} \right )} & P\sim P_{m} 
	\end{array}
	\right. \ .
\end{equation}
Note in particular that the leading small momentum behavior of $\xi_{cr}^{ir}$ does correspond to the $AdS_5\times S^5$ scenario.

The asymptotic form for the length function are more subtle to write. We first discuss the small momentum case. We approximate the integrand in the region of both limits of integration, and then compute the definite integral using primitives for our two approximations and an appropriate ``constant of integration'' (which is really a function of momentum). More specifically, we rewrite the length function as 
\begin{equation}
	l_{\xi_0} = 2\,\sqrt{\alpha'} G^{1/4} \left| \int_{\xi_0}^{\xi_{crit}} d\xi\ \frac{1}{\xi^2} \left ( \frac{\xi}{\sqrt{1+\xi^6}} - P^2 \right )^{-1/2} \right|\ ,
\end{equation}
and then use a first order expansion of the square root in $P^2$ for the integrand near the $\xi_0$ limit. For the other limit, we define 
\begin{equation}
	\Psi \equiv \frac{\xi}{\sqrt{1+\xi^6}} - P^2 
\end{equation}
and observe that the $\Psi \rightarrow 0$ limit is exactly the $\xi_{cr}$ limit. Approximating $\Psi$ in the IR case of large $\xi$ by $\xi^{-2} - P^2$ and in the UV case by $\xi - P^2$, we find the small momentum limits for $l^{ir}$ and $l^{uv}$.

For the large momentum limit, we encounter the problem of the integrand diverging while the interval of integration vanishes. Thus, we perform a rescaling 
\begin{eqnarray}
	{\xi} & = & {\xi_0 \left ( 1 + \lambda \sqrt{1-\frac{P}{P_{m}}} \right )} \\
	{P} & = & {P_{m} (1 - \epsilon^2 )} 
\end{eqnarray}
which changes our parameters from $P$ and $\xi$ to $\epsilon$, which is small, and $\lambda$, which ranges approximately from $0$ to $1$ for IR and $-1$ to $0$ for UV. Neglecting terms which are of order $\epsilon^2$, the integrand becomes 
\begin{equation}
	\frac{\left(1+\xi ^6\right)^{1/4}}{\xi^2 \left(\xi -P^2 \sqrt{1+\xi^6}\right)^{1/2}} \simeq 2^{-1/2} 3^{1/4} \left ( 1 - \frac{3}{2} \epsilon \lambda \right ) \left ( 1 - \left ( \lambda - \frac{\epsilon}{6} \right )^2 \right )^{-1/2} \left ( 1 - \frac{\frac{2\epsilon\lambda}{3}(1-\lambda^2)}{1 - \lambda^2 + \epsilon \lambda - \frac{2}{3} \epsilon \lambda^3} \right )^{1/2} 
\end{equation}
and the corrective term is well approximated by 
\begin{equation}
	\frac{\frac{2\epsilon\lambda}{3}(1-\lambda^2)}{1 - \lambda^2 + \epsilon \lambda - \frac{2}{3} \epsilon \lambda^3} \simeq \frac{2\lambda\epsilon}{3}\ . 
\end{equation}
Using these simplifications, we can analytically integrate, finding the large momentum asymptotes.

\begin{figure}
	\begin{center}
		\includegraphics[width=7in]{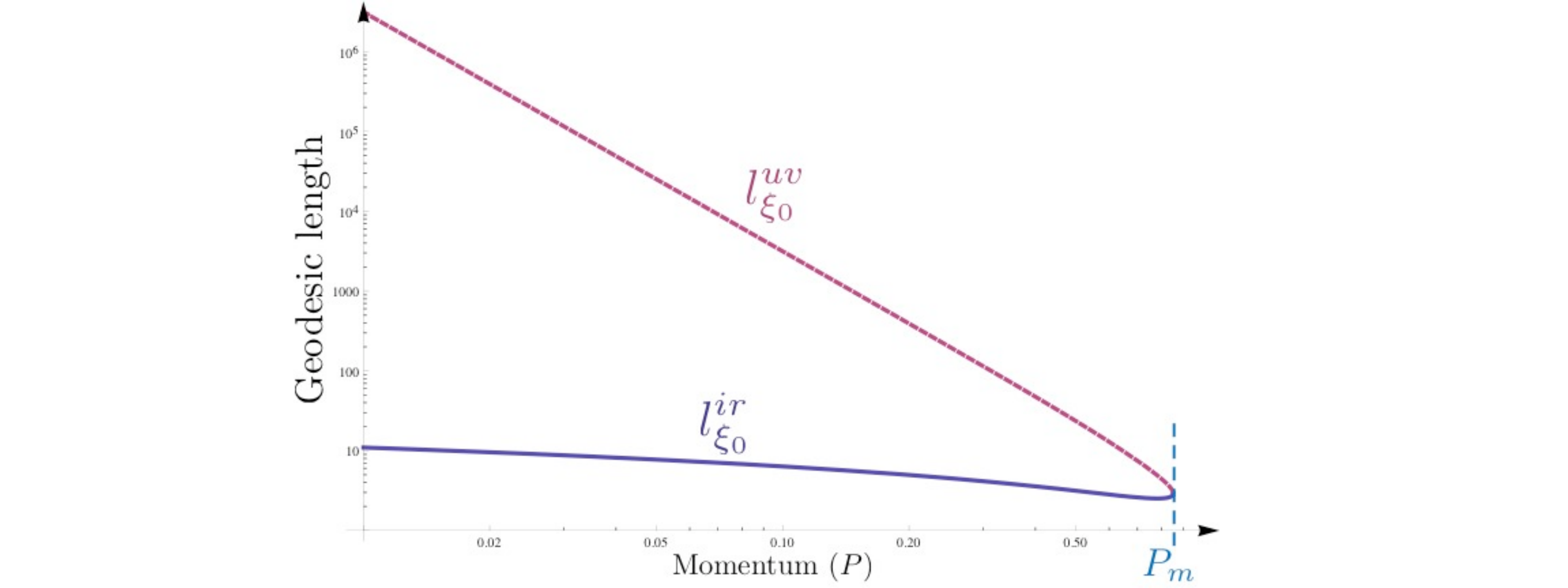}
	\end{center}
	\caption{Length of spacelike geodesics as a function of momentum $P$. Note that at maximum momentum $P_m$ the length is finite and given by $l_{cr}$.} \label{fig:lengthplots} 
\end{figure}

The asymptotes in both regimes for IR geodesics to leading order are 
\begin{equation}
	l^{ir}_{\xi_0}=\sqrt{\alpha'} G^{1/4} \left\{ 
	\begin{array}{ll}
		-2 \ln P + \text{const.} & P\ll P_{m} \\
		\ell_{cr} -5 P_{m}^3\sqrt{1-\frac{P}{P_{m}}} & P\sim P_{m} 
	\end{array}
	\right. 
\end{equation}
and for UV geodesics, they are 
\begin{equation}
	l^{uv}_{\xi_0}=\sqrt{\alpha'} G^{1/4}\left\{ 
	\begin{array}{ll}
		\frac{\pi }{P^3}+\text{const.} & P\ll P_{m} \\
		\ell_{cr}+5 P_{m}^3 \sqrt{1-\frac{P}{P_{m}}} & P\sim P_{m} 
	\end{array}
	\right. \ ,
\end{equation}
where
\begin{equation}
	l_{cr}=\frac{\pi}{2^{1/3} P_m}\ .
\end{equation}
Figure~\ref{fig:lengthplots} shows a plot of the two length functions as a function of momentum $P$.
Note that the small momentum limit $l^{ir} \sim -2 \log P$ is what we expect from the IR limit. This corresponds to a correlation function of the form 
\begin{equation}\label{eq:localcorr}
	C_1^{\Lambda_m}\sim P^{2 M_0 G^{1/4}} \ \ \ \ \ \ \mbox{Small momentum}
\end{equation}
as expected.
On the other hand, at large momenta we have
\begin{equation}
	C_1^{\Lambda_m}\sim e^{5 M_0 G^{1/4} P_{m}^3 \sqrt{1-\frac{P}{P_{m}}}}  \ \ \ \ \ \ \mbox{Large momentum.}
\end{equation}

\begin{figure}
	\begin{center}
		\includegraphics[width=7in]{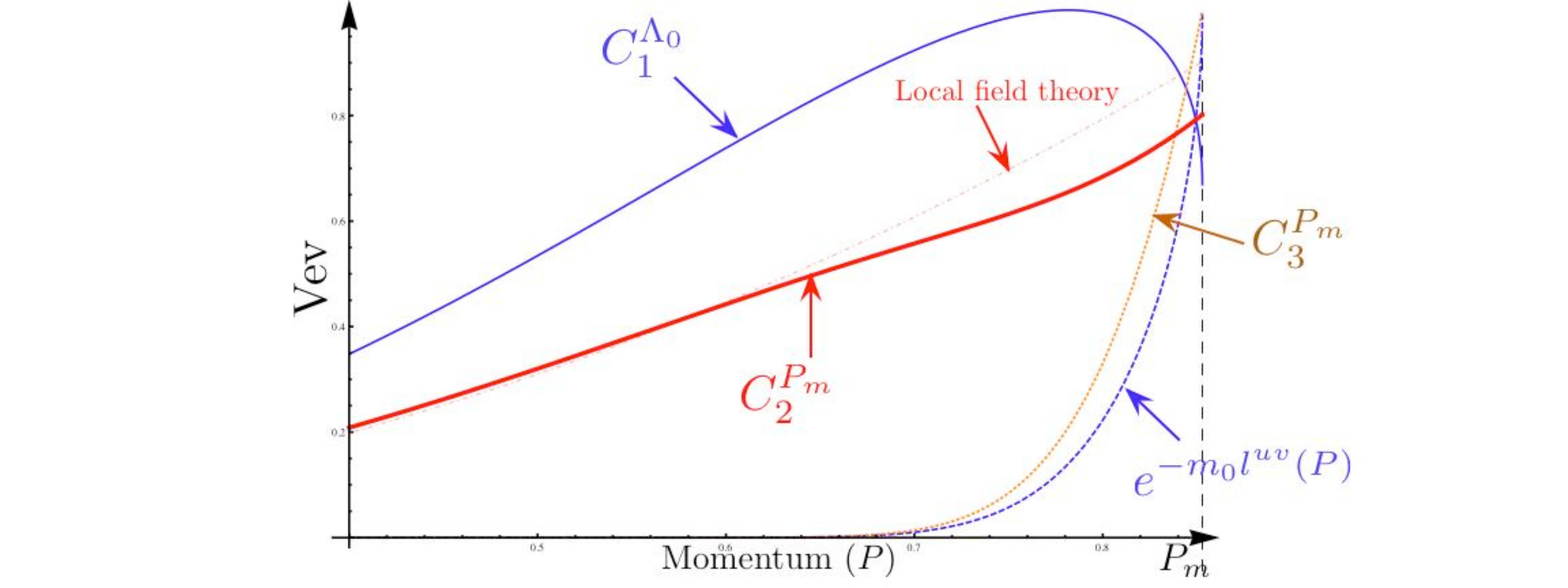}
	\end{center}
	\caption{Various correlation functions as a function of $P$. Note that there is a finite value for the correlators at $P=P_m$.} \label{fig:correlationplots} 
\end{figure}

With $l^{uv}$ one gets instead the asymptotes
\begin{equation}
	e^{-m_0 l^{uv}}\sim\left\{ 
	\begin{array}{ll}
		e^{-M_0 G^{1/4}\frac{\pi }{P^3}} & P\ll P_{m} \\
		e^{-5 M_0 G^{1/4} P_{m}^3 \sqrt{1-\frac{P}{P_{m}}}} & P\sim P_{m} 
	\end{array}
	\right. \ .
\end{equation}
For the combination $e^{-m_0 l^{ir}}+e^{-m_0 l^{uv}}$, we have $e^{-m_0 l^{ir}}\gg e^{-m_0 l^{uv}}$ for $P\ll P_{m}$. So, we could use~(\ref{eq:localcorr}) once again
\begin{equation}
	C_2^{P_m}\sim P^{2 M_0 G^{1/4}} \ \ \ \ \ \ \mbox{Small momentum.}
\end{equation}
For $P\sim P_m$, the combination $e^{-m_0 l^{(ir)}}+e^{-m_0 l^{(uv)}}$ would scale as (with $|\alpha|=|\beta|$)
\begin{equation}
	C_2^{P_m}\sim \cosh \left( 5 M_0 G^{1/4} P_{m}^3 \sqrt{1-\frac{P}{P_{m}}}\right) \ \ \ \ \ \ \mbox{Large momentum.}
\end{equation} 
The exponential forms are somewhat reminiscent of (but not identical to) correlators computed in certain other non-local theories~\cite{Rozali:2000np,Gross:2000ba}.
Figure~\ref{fig:correlationplots} shows plots of these various correlators.

\section{An application to cosmology}

A non-local theory with full $SO(3)$ symmetry offers tantalizing prospects for cosmological applications. Perhaps there are qualitative features that may trickle down from the particulars of the PFT case to more general and realistic realizations. In this section, we present a set of speculative yet intriguing conclusions arising from a straightforward embedding of the results of the previous sections into inflationary cosmology.

Let us assume that prior to the inflationary epoch, when temperatures were of
the order of say $10^{15}\mbox{ GeV}$, the primordial plasma involved strongly coupled non-local dynamics akin to the PFT we have studied. Let it also be the case that by this period of cosmological evolution, quantum gravity effects have been tamed and one is dealing with traditional General Relativity with a gravitating matter sector that is described by the PFT. The existence of such an epoch is a generic expectation in many cosmological scenarios. We propose that the qualitative effects of the non-local dynamics can be tucked into initial conditions for the standard equations of evolution for small fluctuations. This is known in the cosmology literature as the minimal approach~\cite{Martin:2003kp,Martin:2004um,Martin:2007bw} (for other recent and interesting realizations of string cosmology, see for example~\cite{Kaloper:2002uj,Borunda:2006fx,Holman:2006ny,Chowdhury:2006pk,Spalinski:2007dv,Barnaby:2007yb,Kallosh:2007ig,Brandenberger:2007bt,Tsujikawa:2003gh,Koh:2007rx,Brandenberger:2007rg}).

\begin{figure}
	\begin{center}
		\includegraphics[width=6in]{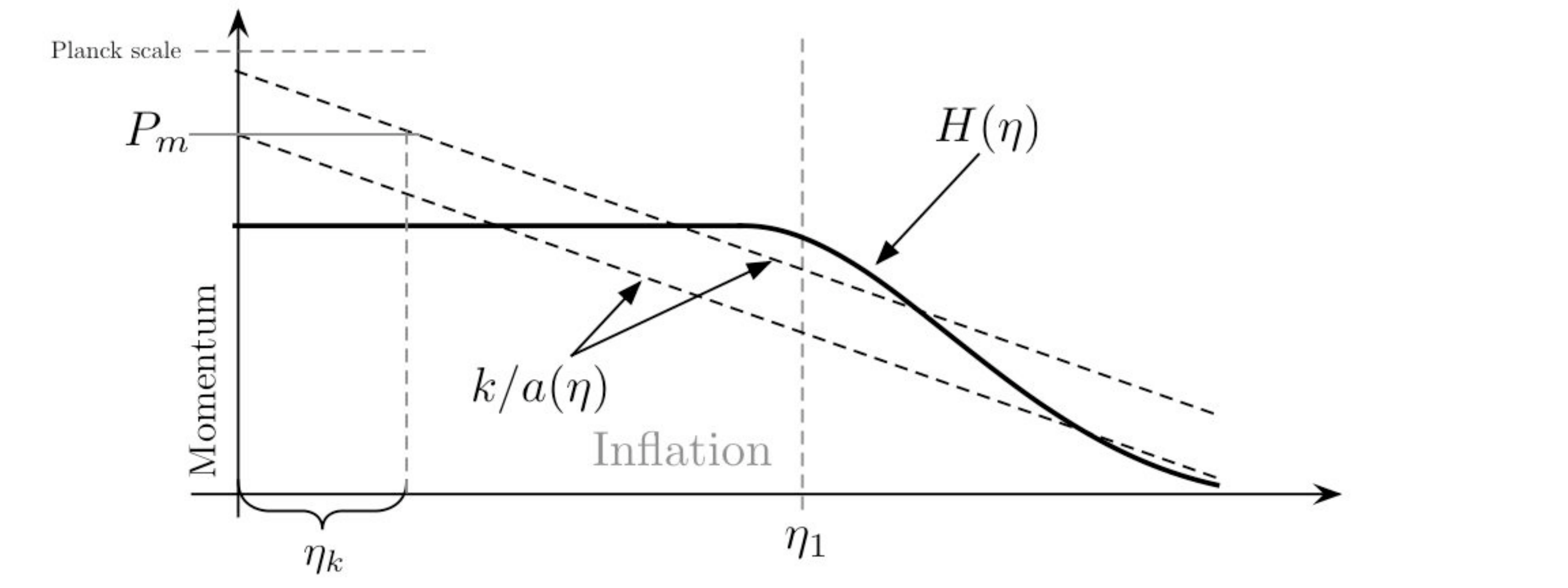}
	\end{center}
	\caption{A rough cartoon of cosmological history through an inflationary epoch. The dotted lines delineate a red-shifted momentum window while the Hubble scale is shown as a solid line.} \label{fig:comsohistory} 
\end{figure}

Hence, at some instant in cosmological time prior to inflation, equal-time correlation functions of matter fluctuations were given by non-local strongly coupled dynamics such as that of the PFT. These initial conditions are then evolved through an inflationary epoch; and the question becomes whether the violent expansion of the universe - which freezes quantum fluctuations - also blows up non-local effects to macroscopic observable length scales. We want to find out what would the signature of non-local dynamics from an earlier time appear as today - within the temperature fluctuations of the CMB (see Figure~\ref{fig:comsohistory}).

\subsection{The minimal approach}

We start with the standard Friedmann-Lema\^{i}tre-Robertson-Walker (FLRW) metric that governs the evolution of the universe through the inflationary epoch (for reviews of modern cosmology, see for example~\cite{Martin:2004um,Mukhanov:1990me,Mukhanov:2003xw,Mukhanov:2005sc})
\begin{equation}
	ds^2=-dt^2+a(t)^2 \sum_i dx_i^2=a^2(\eta) \left( -d\eta^2+\sum_i dx_i^2\right) 
\end{equation}
where $-\infty<\eta\leq 0$ is the cosmological time. There are various scenarios for inflationary expansion, from slow-roll to hybrid to deSitter inflation. Generically, the scale factor is given by 
\begin{equation}
	a(\eta)=\frac{(-\eta)^\nu}{(-\eta_0)^\nu} 
\end{equation}
where for example $\nu=-1$ for deSitter exponential expansion and $\nu=-1-\epsilon$ for slow roll. We have also normalized the scale factor so that it is one at the present epoch $\eta=\eta_1$.

Consider matter field fluctuations $\mu_k(\eta)$ of momentum mode $k$ in this background geometry as qualitative representatives of scalar fluctuations in the primordial plasma. Such small fluctuations would satisfy the equation of motion 
\begin{equation}
	\label{eq:eom} {\mu_k}''+\omega(\eta) \mu_k=0 
\end{equation}
where the frequency is time dependent 
\begin{equation}
	\omega(\eta)=\sqrt{k^2 - \frac{a''}{a}}=k\,\sqrt{1-\frac{\nu (\nu-1)}{(k \eta)^2}} \ .
\end{equation}
Hence the modes will be subject to the Schwinger effect~\cite{Martin:2007bw}. 

The solution to equation~(\ref{eq:eom}) is given by 
\begin{equation}
	\mu_k(\eta)=\alpha_k\, \mu^{(0)}_k(\eta)+ \beta_k\, \mu^{(0)*}_k(\eta) 
\end{equation}
where 
\begin{equation}
	\mu^{(0)}_k(\eta)=-\frac{2\pi\, i}{m_P} e^{-i\,\pi\,\nu/2} \sqrt{-\eta}\, H^{(1)}_{\frac{1}{2}-\nu}(-k\eta) \ ,
\end{equation}
with $H^{(1)}_{\alpha}$ being the Hankel function of the first kind, and $\alpha_k$ and $\beta_k$ being complex constants to be fixed by initial conditions - {\em i.e.} they are the Bogoliubov coefficients. 

The needed asymptotic forms of $\mu_k^{(0)}(\eta)$ are 
\begin{equation}
	\mu^{(0)}_k(\eta)\simeq \left\{ 
	\begin{array}{ll}
		\frac{2^{\frac{3}{2}-\nu} e^{-\frac{3}{2} i\,\pi\,\nu} \Gamma\left(\frac{1}{2}-\nu \right)}{m_P\,\sqrt{k}}(k \eta )^{\nu} & \mbox{For contemporary times $\eta\rightarrow 0$} \\
		\frac{\sqrt{2 \pi }}{m_P} \frac{e^{-i\,k\,\eta} (i (\nu -1) \nu -2 k \eta )}{\sqrt{k} (k\,\eta) } & \mbox{For early times $\eta\rightarrow -\infty$} 
	\end{array}
	\right. \ .
\end{equation}

At some early moment in time $\eta_k\rightarrow -\infty$, the frequency is given by 
\begin{equation}
	\label{eq:freq} \omega(\eta_k)=k \sqrt{1-\frac{\nu(\nu-1)}{\eta_k^2 k^2}}\simeq k \ .
\end{equation}
At this time, let the red-shifted momentum $k$ be
\begin{equation}
	\frac{k}{a(\eta_k)}\simeq \frac{k (-\eta_0)^{\nu}}{(-\eta_k)^{\nu}}\sim \frac{1}{G^{1/3}\Delta} \Rightarrow
	\sigma_k\equiv -\frac{1}{k \eta_k}=
	-\frac{H(\eta_0)}{\nu} \frac{1}{k^{1+\frac{1}{\nu}} (G^{1/3} \Delta)^{1/\nu}} \label{eq:benchmark}
\end{equation}
where the Hubble constant is given by $H(\eta)=\nu/(a(\eta) \eta)$.
We want to impose initial conditions onto~(\ref{eq:eom}) at this moment in time $\eta_k$. Given that the maximum momentum is within reach as seen from~(\ref{eq:benchmark}), we need to set the UV cutoff at the throat $\xi_0$ in the PFT.

We write the initial conditions of the fluctuations as 
\begin{equation}
	\mu_k(\eta_k)= -\frac{r_{1}^k e^{i\, \varphi_{1}^k}}{\sqrt{2 \omega(\eta_k)}} \frac{4\sqrt{\pi}}{m_s} 
\end{equation}
and 
\begin{equation}
	\mu'_k(\eta_k)=i\, \sqrt{\frac{\omega(\eta_k)}{2}} \frac{4\sqrt{\pi}\left(r_{2}^k e^{i\, \varphi_{2}^k}\right)}{m_s} \ ,
\end{equation}
with $r_1^k$, $r_2^k$, $\varphi_1^k$, and $\varphi_2^k$ being real constants parameterizing the boundary conditions, and $m_s$ is the string mass to arrange for the correct units; they must satisfy the constraint 
\begin{equation}
	\label{eq:normalization} r_1^k r_2^k \cos \left( \varphi_1^k-\varphi_2^k\right)=1 
\end{equation}
since one needs $|\alpha_k|^2-|\beta_k|^2=1$. The Bogoliubov coefficients become 
\begin{equation}
	\alpha_k= \frac{e^{i \phi_2^k} e^{-\frac{i}{\sigma_k}} e^{i \Delta \phi_k}}{\sqrt{k} \sqrt{\omega(\eta_k)}} \frac{k\, \sigma_k \left(-\nu ^2-\frac{i (\nu -1) \nu }{\sigma_k }+\nu +\frac{2}{\sigma_k^2}\right) r_1^k-\omega(\eta_k)\left(i (\nu -1) \nu -\frac{2}{\sigma_k }\right) r_2^k }{(\nu -2) (\nu -1) \nu (\nu +1)\sigma_k+\frac{4}{\sigma_k}} 
\end{equation}
\begin{equation}
	\beta_k= \frac{e^{i \phi_2^k} e^{\frac{i}{\sigma_k}} e^{i \Delta \phi_k}}{\sqrt{k} \sqrt{\omega(\eta_k)}} \frac{k\, \sigma_k \left(-\nu ^2+\frac{i (\nu -1) \nu }{\sigma_k }+\nu +\frac{2}{\sigma_k^2}\right) r_1^k+\omega(\eta_k)\left(-i (\nu -1) \nu -\frac{2}{\sigma_k }\right) r_2^k }{(\nu -2) (\nu -1) \nu (\nu +1)\sigma_k+\frac{4}{\sigma_k}} \ .
\end{equation}

The power spectrum of scalar perturbations - which would be proportional to temperature fluctuations in the 
CMB~\cite{Martin:2003kp} - would be seen today as 
\begin{equation}
	\label{eq:ps} k^3 \mathcal{P}(k)=\left.\frac{2\,k^3}{\pi^2} \frac{|\mu_k(\eta)|^2}{a^2(\eta)}\right|_{\eta\rightarrow \eta_0} \ .
\end{equation}
Putting things together, we get 
\begin{eqnarray}\label{eq:theP}
	k^3 \mathcal{P}(k)&=& 
	\frac{2^{3-2\nu} (H(\eta_0))^{-2 \nu}
	   (-\nu)^{2 \nu } \Gamma\left(\frac{1}{2}-\nu \right)^2}{m_s^2 \pi^2} k^{2 (\nu +1)} \left[(r_1^k)^2+(r_2^k)^2\right] \times \nonumber \\
	   & &\left[\frac{\left((r_1^k)^2-(r_2^k)^2\right) \cos \left(\pi  \nu
	   +\frac{2}{\sigma_k }\right)-2 \sqrt{(r_1^k)^2 (r_2^k)^2-1} 
	\sin \left(\pi\nu +\frac{2}{\sigma_k}\right)}{(r_1^k)^2+(r_2^k)^2}+1\right]
\end{eqnarray}
where we have expanded in $\sigma_k$ to leading order, dropping a series of positive powers of $1/\sigma_k$. Using equation~(\ref{eq:normalization}), we are left with two free sets of modes to fix; say $r_1^k$ and $r^k_2$; this is because, as we see from the previous equation, the power spectrum does not depend on the combination $\phi_1^k+\phi_2^k$ (this is true to all orders in $\sigma_k$).

To fix $r_1^k$ and $r_2^k$, we assume that the initial vacuum is in a configuration that saturates the uncertainty bound 
\begin{equation}
	\left|\left< \Pi^\dagger \Pi\right>\right|^{1/2} \left|\left< \mathcal{O}^\dagger \mathcal{O}\right>\right|^{1/2} \simeq 1 
\end{equation}
where $\Pi$ and $\mathcal{O}$ are a canonical operator pair. We then have 
\begin{equation}
	|\mu'_k(\eta_k)|=k |\mu_k(\eta_k)| \Rightarrow r_2^k=r_1^k\frac{k}{\omega(\eta_k)}\ .
\end{equation}
We contrast this with the Danielsson vaccum~\cite{Danielsson:2002kx}, where one would impose the condition $\mu'_k(\eta_k)=k \mu_k(\eta_k)$. The uncertainty relation does not require such a stringent statement; the freedom in the additional phase is, as we shall see, where the information about the non-local dynamics gets encoded.

Hence, we are left with one set of modes to determine; and this data is to come from the PFT correlators. 
We expect that $r_1^k\sim \sqrt{k} |\mu_k|$ relates to $\left< \mathcal{O}^\dagger(P) \mathcal{O}(P)\right>_{P_{m}}^{1/2}=\sqrt{C_2^{P_m}(P)}$. To determine the proportionality constant, we note that the IR limit of the PFT is scale invariant, being dual to $AdS_5\times S^5$; and in this regime the correlator is given by $P^{2 M_0 G^{1/4}}$. The normalization of $|\mu_k|$ on the other hand is such that scale invariance for $\nu=-1$ implies $r_1^k\sim 1$, independent of $k$. This tells us that we have
\begin{equation}
	r_1^k\sim \frac{\sqrt{C_2^{P_m}(P)}}{P^{M_0 G^{1/4}}}\ .
\end{equation}
Note that the momentum $P$ is $k$ rescaled as in $P \sim G^{1/3} \Delta\, k$. Hence, these steps fix the power spectrum across all scales $k$, bringing in the non-local effects at high momenta. 

\begin{figure}
	\begin{center}
		\includegraphics[width=6in]{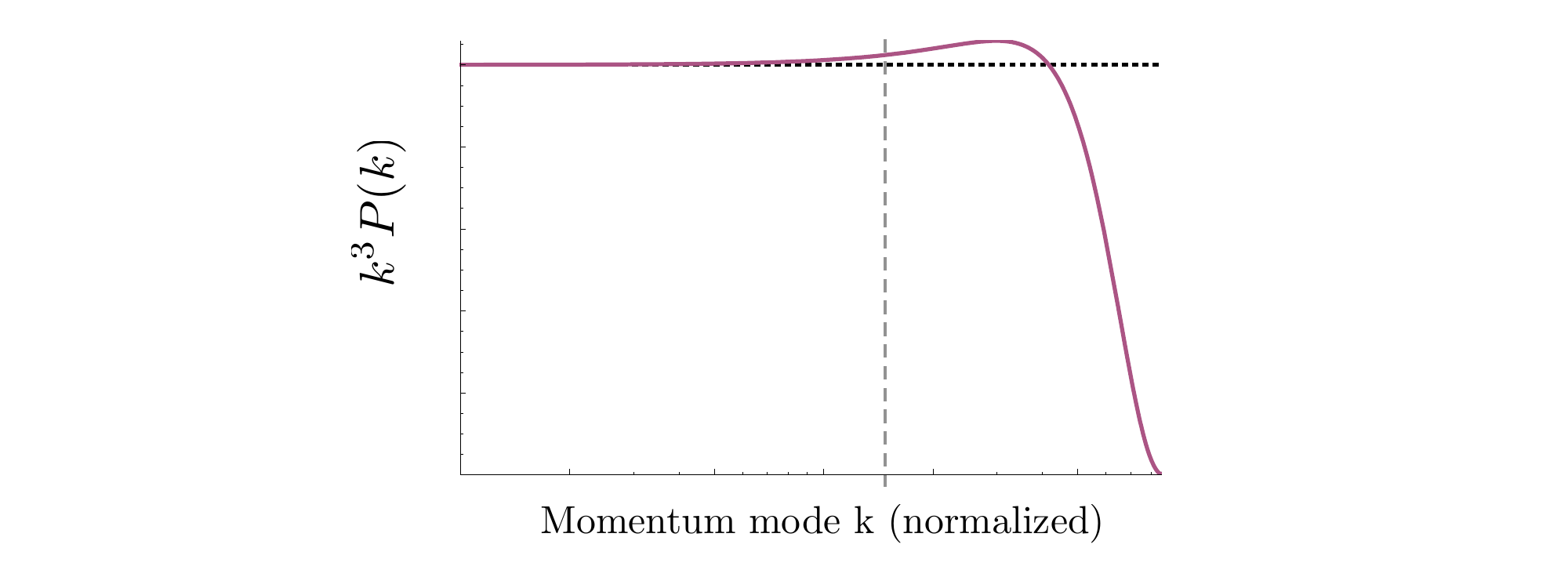}
	\end{center}
	\caption{A semi-log plot of the normalized power spectrum with initial conditions set by the correlators of the PFT. We have chosen $\nu=-1$ (deSitter inflation), $G=10^{24}$, and $\Delta=10^{-2}$. The horizontal dotted line corresponds to a scale invariant spectrum. The vertical dotted line corresponds to $k\sim G^{-1/3}\Delta^{-1}$.} \label{fig:cosmoplot} 
\end{figure}

Putting everything together, Figure~\ref{fig:cosmoplot} shows a plot of the power spectrum for an interesting set of parameters. We see from the Figure that the scale invariant profile gets gradually corrected as we probe wavenumbers of the order $G^{-1/3} \Delta^{-1}$, the minimal length scale of the PFT. 

\section{Discussion}

In this work we have explored a new mechanism for implementing non-local dynamics in string theory; one that involves quantum entanglement of UV and IR physics, and a minimal length scale enhanced at strong coupling. We also applied the results to structure formation in cosmology with intriguing conclusions that suggest improved prospects for finding stringy signature in the CMB.

It would be interesting to explore further the new aspects of the holographic dictionary that arise in the presence of an apparent horizon. Having the throat in the geometry associated with the UV completion of the dual theory is suggestive of the original premise of gravitational holography: the decoupling limit focuses onto the throat of a geometry that otherwise asymptotes to Minkowski space; it appears that in the PFT setting, after the decoupling limit is taken, one has yet another throat within reach. But now we are able to extend the holographic dictionary across this benchmark since all of the space is available for holography. Perhaps the lessons from the PFT can help understand holography in asymptotically Minkowski spacetimes as well.

The PFT we used in the current work is one of several possible realizations of the D3 branes in a Melvin universe scenario. In particular, one can realize other PFTs with less or no supersymmetry that may be particularly interesting. The IR limit of all of these would be $\mathcal{N}=4$ SYM. Hence there is a rich landscape in renormalization group flow that one can attempt to map. Perhaps the strong coupling analysis would then help in identifying deformations of the $\mathcal{N}=4$ SYM Lagrangian towards these PFTs.

Our analysis was confined to using spacelike geodesics. One can also consider timelike geodesics for the computation of more general correlators. The geodesic analysis has helped in mapping some of the intricate and new aspects of the holographic dictionary; one can now also attempt to understand supergravity field propagation in the bulk so as to move beyond the optical approximation. In particular, one would be interested in expanding the field dynamics about the throat instead of the asymptotic boundaries. Other restrictions on our analysis discussed in Section 2 can also be circumvented by duality transformations of the bulk space. For example, small $\phi$ circle would lead to the T-dual geometry which will necessarily exhibit new dynamics for geodesics.

In the cosmology application, there are various new directions to take. One different realization would involve imposing the initial conditions at some fixed time $\eta_0$ for all modes. This would pick the correlators with fixed energy cutoff  instead - given that an argument for a fixed initial time for new physics is in the spirit of using the red-shifted temperature as the benchmark. One can also easily develop the power spectrum for tensor fluctuations. All these are rather straightforward generalizations; we do not however expect that they lead to any new qualitative features given the general structural form of the power spectrum given in~(\ref{eq:theP}).

Finally, it would be interesting to better understand the physical role played by the product structure of the Fock space we employed in making sense of the UV-IR mixing. Perhaps this is more than a computational tool but instead is hinting at a more general formulation of the PFT. 

\section{Acknowledgments}

We would like to thank Sam Skillman for contributing to the initial stages of this work. This research was supported by Research Corporation Grant No. CC6483.

\vspace{1in}
{\huge \bf Appendices}

\vspace{0.5in}
{\Large \bf Appendix A: Degrees of freedom}
\vspace{0.25in}

The c-function of a theory may heuristically be defined as a measure of the effective number of degrees of freedom as a function of energy scale. In two dimensions, this notion is well defined and the c-function matches onto the central charge of the conformal field theories that typically arise as UV or IR fixed points of renormalization group flow~\cite{Zamolodchikov:1986gt}. In more than two dimensions, the notion of a c-function of a local theory is less developed; for non-local dynamics, the concept is even less transparent. In the context of the dictionary of gravitational holography, a natural geometric quantity arises in the bulk for the c-function of the dual theory~\cite{Sahakian:1999bd}. The c-function at strong coupling would be written as
\begin{equation}\label{eq:cfunctiondef}
	c\propto \frac{1}{G_D \sqrt{h} |\Theta|^{D-2}} \ ,
\end{equation}
where $G_D$ is the gravitational constant of the D dimensional bulk, $h$ is the determinant of the metric projected from the bulk onto the $D-1$ dimensional boundary, and $\Theta$ is the rate of expansion/contraction of a congruence of null geodesics projected from the boundary into the bulk. We apply this proposal to the PFT; and we also analyze two other cases to contrast the conclusions: $3+1$ dimensional Non-Commutative Super Yang-Mills (NCSYM) theory and $1+1$ dimensional Non-Commutative Open String (NCOS) theory.

\vspace{0.5in}
{\Large {\bf A.1:} PFT}
\vspace{0.25in}

\begin{figure}
	\begin{center}
		\includegraphics[width=6in]{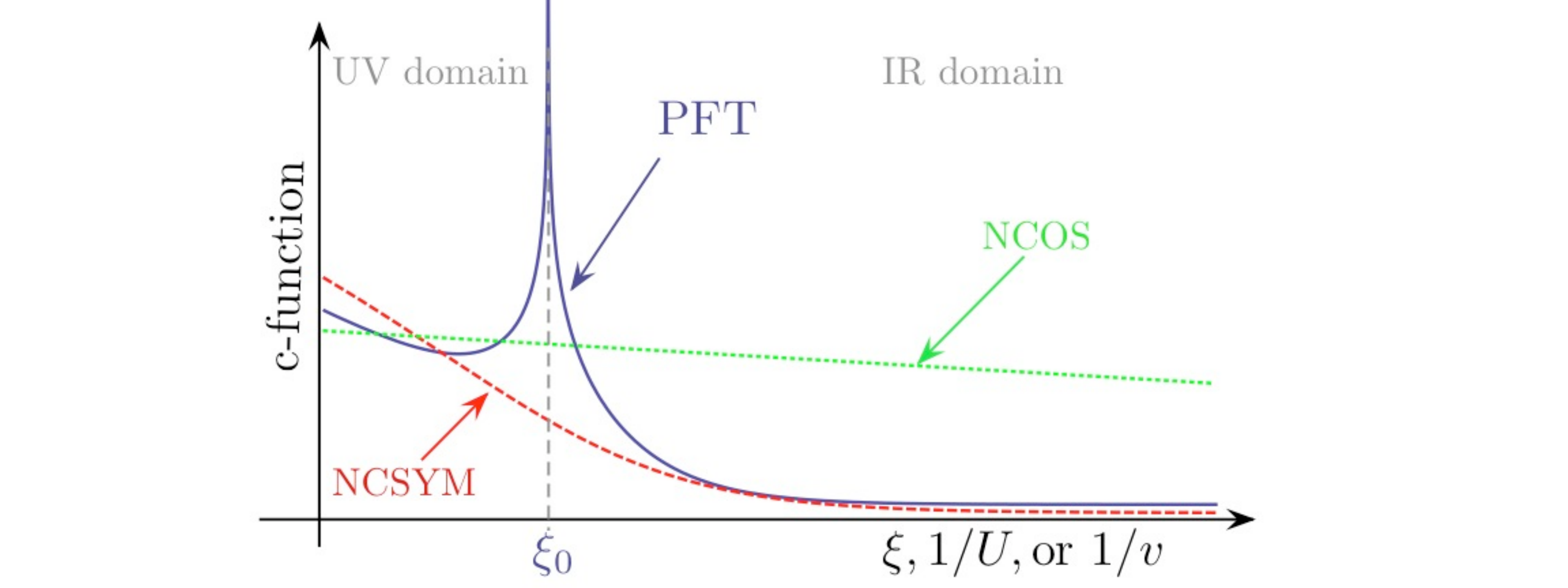}
	\end{center}
	\caption{C-functions for the various theories considered in this Appendix.} \label{fig:puffcfunc} 
\end{figure}

Using~(\ref{eq:thetheta}) and~(\ref{eq:cfunctiondef}), we get a candidate for a c-function of the PFT at strong coupling  
\begin{equation}\label{eq:cpuff}
	c\propto \frac{N^4}{G^{2} \Delta}\frac{\left( 1+\xi^6\right)^{9/2}}{\xi^9|\xi^6-\xi_0^6|^3} 
\end{equation}
where we used the relation 
\begin{equation}
	G_5=\frac{G_{10}}{\mbox{Vol}5}\simeq \frac{g_s^2 {\alpha'}^4}{{\alpha'}^{5/2} (4\pi N)^{5/4}\Delta^{-1} (1+\xi^6)^{3/4} \xi^{-9/2}} \ .
\end{equation}
Given the geodesics used in this computation carry R-charge, we should perhaps interpret this as a measure of non-local degrees of freedom of given R-charge. Figure~\ref{fig:puffcfunc} shows a plot of this c-function. We see that it diverges at the throat $\xi_0$ suggesting the UV completion of the theory sits at $\xi=\xi_0$. A diverging number of degrees of freedom from the perspective of the IR regime suggests that the UV completion is a non-local theory akin to a string theory - with an infinite tower of excitations. However, there is another twist to the story: it is believed that c-functions in higher dimensions may be monotonic and decrease with lower energy scales, in analogy to the two dimensional scenario; in this non-local theory, this function violates both of these properties. While the behavior in the IR domain is consistent with a c-function that is monotonically decreasing with lower energies, the function flips around at $\xi_0$ and again at $\xi=1/\sqrt{2}$ in the UV domain. The first flip is consistent with associating a second theory as the holographic dual to the UV region of the bulk. The flip at $\xi=1/\sqrt{2}$ is however perplexing. 

A c-function would fail the monotonicity condition if the bulk space does not satisfy the null energy condition. We can check this by computing the Ricci tensors
\begin{equation}
	R_{XX}=\frac{\xi ^4 \left(16 \xi ^{12}-58 \xi^6+7\right)}{4 \left(1+\xi^6\right)^3}\ \ \ ,\ \ \ 
	R_{\xi\xi}=\frac{-4 \xi ^{12}+46 \xi ^6-4}{\xi^2\left(1+\xi^6 \right)^2}\ .
\end{equation}
It is then easy to see that $R_{ab} n^a n^b$ fails to be positive for any null geodesics if $\xi<14^{-1/6}$. This does not exactly match with the flip-over point $\xi=1/\sqrt{2}$ because we approximated the effect of compactification to five dimensions by writing $G_5\sim G_{10}/\mbox{Vol}5$ in computing~(\ref{eq:cpuff}). The bottom line is that a new dynamical structure arises for $\xi<\xi_0$ which perhaps invalidates the use of~(\ref{eq:cfunctiondef}) - the latter being designed with local field theories in mind. Yet it seems to work sensibly in the region that asymptotes to $AdS_5\times S^5$ in the IR.
Let us proceed by repeating this analysis for other non-commutative theories that arise in string theory; we shall see that the PFT is rather unusual in this respect. 

\vspace{0.5in}
{\Large {\bf A.2:} $3+1$ dimensional NCSYM}
\vspace{0.25in}

The holographic dual of $3+1$ dimensional NCSYM theory~\cite{Hashimoto:1999ut,Cai:1999aw} is described by the metric
\begin{eqnarray}
	\label{eqn:ncymmetric} {ds_{str}^2} & = & {\alpha' \left [ \frac{U^2}{\sqrt{\lambda}} (-dt^2 + dx_1^2) + \frac{\sqrt{\lambda} U^2}{\lambda + U^4 \Delta^4} ( dx_2^2 + dx_3^2 ) \right .} \\
	\nonumber {} & {} & {\left . \hspace{1cm}+ \frac{\sqrt{\lambda}}{U^2} dU^2 + \sqrt{\lambda} \cdot d\Omega_5^2 \right ]}\ ,
\end{eqnarray}
where $U$ is the holographic coordinate and the UV-IR relation is given by $E\sim U/\sqrt{\lambda}$.
The space also has a nontrivial dilaton
\begin{eqnarray}
	{e^\phi} & = & {\frac{\lambda}{4\pi N} \sqrt{\frac{\lambda}{\lambda + \Delta^4 U^4}}}\ . 
\end{eqnarray}
This NCSYM theory has non-commutativite spatial coordinates, $[x_2, x_3] \ne 0$. 

For spacelike geodesics with both endpoints at fixed $U$, we use the momentum conservation statements and the spacelike geodesic normalization condition, and we find the turn-around points $U'(\tau) = 0$
\begin{eqnarray}
	{\frac{\lambda p_1^2 + (U^4 \Delta^4 + \lambda) (p_2^2 + p_3^2)}{U^2 (U^4 \Delta^4 + \lambda)^{1/4}}} & = & {2 \alpha' \sqrt{\pi N} \lambda^{-1/4}}\ . 
\end{eqnarray}
This equation admits any value of momentum $p_1$ in the $x_1$ direction; however, there is a maximum momentum in the noncommuting directions at
\begin{eqnarray}
	{U_{{crit}}} & = & {\frac{(2\lambda)^{1/4}}{\Delta}}\ . 
\end{eqnarray}
So far, NCSYM dynamics qualitatively agrees with PFT case in that it exhibits a maximum momentum for equal time operator insertions; although with the exception that it breaks the $SO(3)$ symmetry of the PFT. 

\begin{figure}
	\begin{center}
		\includegraphics[width=6in]{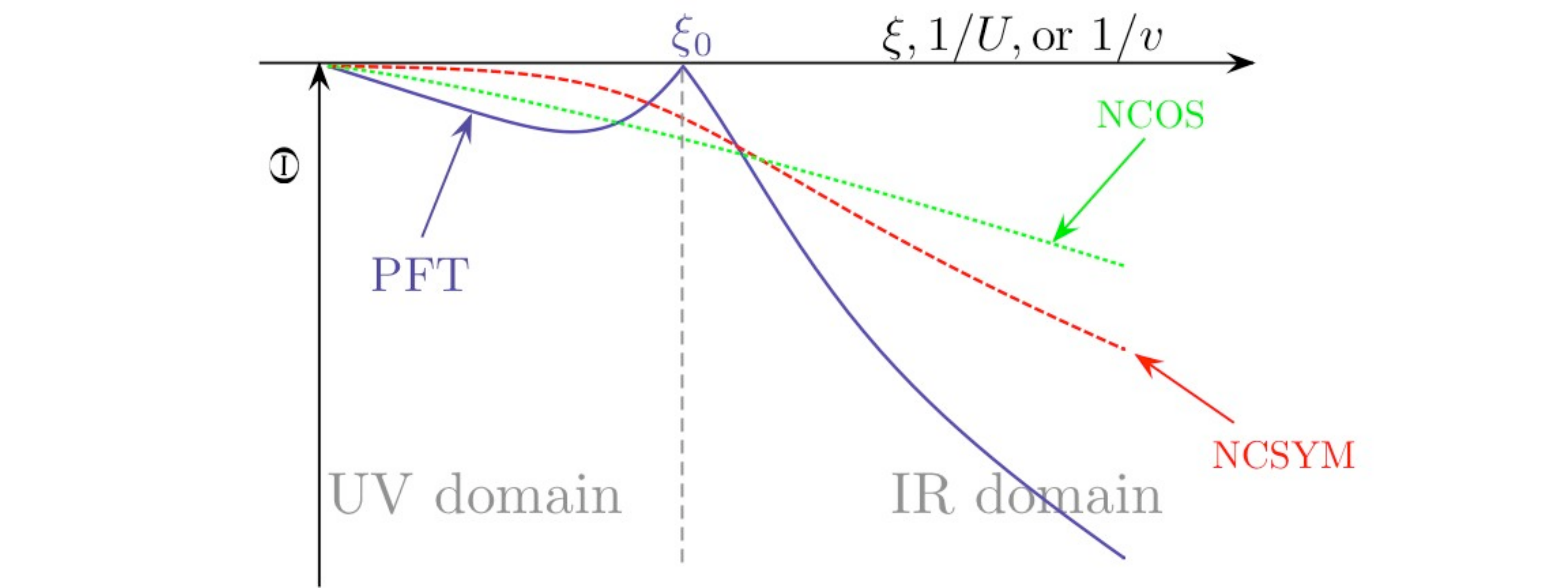}
	\end{center}
	\caption{$\Theta$ for the various theories considered in this Appendix.} \label{fig:thetaplots} 
\end{figure}

The tangent to null geodesics in this space is given by
\begin{equation}
	{u^a} =  {\frac{\lambda^{5/4}}{2 \alpha' \sqrt{N \pi} U^2 (U^4 \Delta^4 + \lambda)^{1/4}}\,}\partial_t {\hspace{0.1cm} \pm\frac{\lambda^{3/4}}{2 \alpha' \sqrt{N \pi} (U^4 \Delta^4 + \lambda)^{1/4}}\, \partial_U} 
\end{equation}
which leads to the expansion/contraction rate
\begin{eqnarray}
	{\Theta} & = & \pm \frac{\lambda^{3/4}}{4 \alpha' \sqrt{N \pi}}{\frac{U^4 \Delta^4 + 6 \lambda}{U(U^4 \Delta^4 + \lambda)^{5/4}}}\ .
\end{eqnarray}
Thus, we find that we always have $\Theta < 0$ as we move towards smaller $U$ or lower energies as seen in Figure~\ref{fig:thetaplots}; according to the Bousso criterion, the holographic screen should be placed at $U = \infty$.
The associated c-function that follows from~(\ref{eq:cfunctiondef}) is 
\begin{equation}
	c\sim \frac{\left(U^4+1\right)^5}{\left(U^4+6\right)^3}\ ,
\end{equation}
with a plot in Figure \ref{fig:puffcfunc}. It smoothly decreases away from the boundary at infinity and is monotonically decreasing towards the IR as expected. We then see that the apparent horizon that arises in the PFT - and the associated flipping of the c-function in the UV - do not necessarily extend to the NCSYM theory. To further explore this contrast, let us next look at NCOS theory. The latter is of a different character as it involves a full open string sector.

\vspace{0.5in}
{\Large {\bf A.3:} $1+1$ dimensional NCOS}
\vspace{0.25in}

$1+1$ dimensional NCOS theory~\cite{Harmark:2000wv,Sahakian:2001xz} has a holographically dual geometry described by the metric
\begin{eqnarray}
	\label{eqn:ncosmetric} {ds_{str}^2} & = & {\Omega^2 \left [ \frac{v^2}{8 \pi^2 \alpha_e} (-dt^2 + \Sigma^2 dy^2 ) + \frac{dv^2}{v^2} + 4 \cdot d\Omega_7^2 \right ]} 
\end{eqnarray}
where
\begin{eqnarray}
	{\Omega^2} & = & {8 \pi^2 \alpha' \frac{\sqrt{G}}{\sqrt{v}} \sqrt{1 + Gv^3}}\ .
\end{eqnarray}
The space also has a nontrivial dilaton
\begin{eqnarray}
	{e^\phi} & = & {\left ( 32 \pi^2 \right )^2 \frac{G^{3/2}}{M} \frac{1 + Gv^3}{v^{3/2}}} \ .
\end{eqnarray}
The holographic direction is labeled by the coordinate $v$ and the UV-IR relation goes as $E\sim v/\sqrt{\alpha_e}$.
NCOS theory has its time coordinate non-commuting with its only spatial coordinate $[t, y] = 2\pi i \alpha_e$. This property does however preserve the full $SO(1,1)$ Lorentz symmetry.

Consider spacelike geodesics with both endpoints at fixed $v$. Using the normalization condition for spacelike geodesics and the conservation of momentum along $y$, we get a condition for the turn-around point $v'(\tau) = 0$ 
\begin{eqnarray}
	{\left ( 32 \pi^2 \alpha_e G^{1/4} \right ) p^2} & = & {\left ( M^{1/2} \alpha' \Sigma^2 \right ) v^{9/4}}\ . 
\end{eqnarray}
Thus, all momenta are allowed and all points are candidates for critical values. If we look at the second-order geodesic equation for $v$, then (setting $t' = 0$) we find
\begin{eqnarray}
	{v''} & = & {\frac{9 v^3 \Sigma^2 (y')^2}{64 \pi^2 \alpha_e} + \frac{7(v')^2}{8v}}\ . 
\end{eqnarray}
Thus, any critical point must be a minimum for $v(\tau)$. This suggests that the UV completion of the NCOS theory sits at $v = \infty$.

Null geodesics in the $(t, v)$ plane have tangent
\begin{eqnarray}
	{u^a} & = & \frac{32 \pi^2 \alpha_e G^{1/4}}{v^{9/4} \alpha' \sqrt{M}}\partial_t \pm \frac{8\sqrt{2} \pi G^{1/4} \sqrt{\alpha_e}}{\alpha' v^{1/4} \sqrt{M}}\partial_v
\end{eqnarray}
and expansion/contraction rate
\begin{eqnarray}
	{\Theta} & = & \pm {\frac{9\sqrt{2} \pi}{\alpha'}} \frac{\sqrt{\alpha_e}}{\sqrt{M}} \frac{ G^{1/4} }{ v^{5/4} }\ .
\end{eqnarray}
Once again, we always have $\Theta<0$ for decreasing $v$ suggesting a holographic screen at $v\rightarrow \infty$, as shown in Figure~\ref{fig:thetaplots}.
We can also compute the candidate c-function along~(\ref{eq:cfunctiondef})
\begin{eqnarray}
	\label{eqn:cfuncncos} {c} & \sim & v\ . 
\end{eqnarray}
Hence, both the c-function and energy decrease with decreasing $v$ (see Figure~\ref{fig:puffcfunc}) as expected. Once again, we see that the PFT stands out in its level of complexity and richness even amongst other non-commutative theories.

\vspace{0.5in}
{\Large \bf Appendix B: Momentum versus size}
\vspace{0.25in}

In the same way as we wrote geodesic length as a function of momentum,
the relation linking displacement $\Delta X$ to momentum is given by 
\begin{equation}
	\Delta X^{uv,ir}= 2\,\left| \int_{\xi_0}^{\xi_{cr}} d\xi\frac{P (1+\xi^6)^{3/4}}{\xi^3 (\xi-P^2 \sqrt{1+\xi^6})^{1/2}} \right| ,
\end{equation}
where the UV/IR label distinguishes between the two geodesics with different $\xi_{cr}$ turning points. We again are forced to treat this integral numerically and through asymptotic forms.

\begin{figure}
	\begin{center}
		\includegraphics[width=6in]{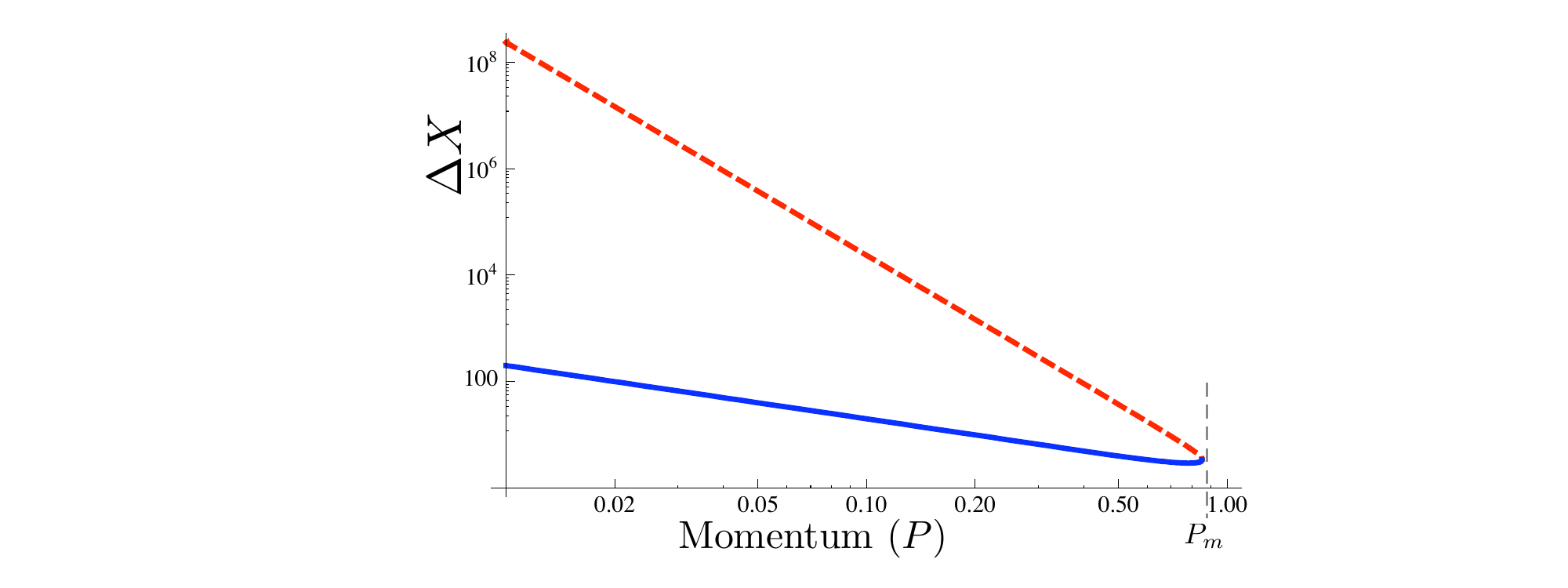}
	\end{center}
	\caption{Displacement $\Delta X$ of spacelike geodesics as a function of momentum $P$.} \label{fig:fourierplot} 
\end{figure}

Using the same techniques as in Section \ref{sec:detailed}, we find
\begin{equation} \label{eq:deltaxir}
	\Delta X^{(ir)}(P)\simeq\left\{ 
	\begin{array}{ll}
		2 P^{-1} & P\ll P_{m} \\
		\frac{\ell_{cr}}{P_{m}} - 5 P_{m}^2 \sqrt{1-\frac{P}{{P_{m}}}} & P\sim P_{m} 
	\end{array}
	\right. 
\end{equation}
and 
\begin{equation} \label{eq:deltaxuv}
	\Delta X^{(uv)}(P)\simeq\left\{ 
	\begin{array}{ll}
		\frac{3\pi}{4} P^{-4} & P\ll P_{m} \\
		\frac{\ell_{cr}}{P_{m}} + 5 P_{m}^2 \sqrt{1-\frac{P}{{P_{m}}}} & P\sim P_{m} 
	\end{array}
	\right. , 
\end{equation}
where
\begin{equation}
	\ell_{cr}=\frac{\pi}{2^{/3} P_m}\ .
\end{equation}
The small momentum behavior $\Delta X^{(ir)} \sim 2P^{-1}$ is what we expect from a local field theory.
Figure~\ref{fig:fourierplot} shows a plot of both functions.

In the language of Section \ref{sec:uvir}, $\Delta X^{-1} \sim \kappa$
instead of $k$.  
Associating $\xi$ and $P$ through (\ref{eq:puffmomxi}) and using the $P \ll P_m$
asymptotics in (\ref{eq:xicritir}), (\ref{eq:xicrituv}), (\ref{eq:deltaxir}), and (\ref{eq:deltaxuv}),
we find $\Delta X \sim \xi^{-2}$ for $\xi \ll \xi_0$ and $\Delta X \sim \xi$ for $\xi \gg \xi_0$.

If we were to attempt to use this relation to transform into configuration space, we must remember to consider both UV and IR geodesics. A given displacement is associated with \emph{two} momenta. In fact, inverting the displacement/position relation is made even trickier since $\Delta X^{(ir)}(P)$ is not even strictly decreasing.  This means that the extent
of an operator increases with momentum in a certain regime; this has been found
to be a generic feature of non-locality.

\begin{figure}
	\begin{center}
		\begin{tabular}
			{cc} {\framebox{
			\begin{tabular}
				{c} 
				\includegraphics[width=1.6in]{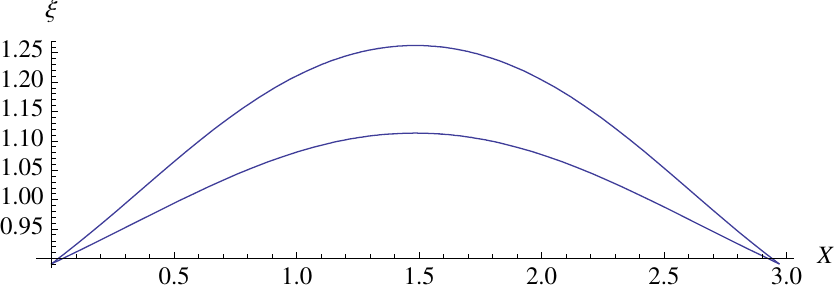} \\
				{$\Delta X = 2.97$} 
			\end{tabular}
			}} & {\framebox{
			\begin{tabular}
				{c} 
				\includegraphics[width=1.6in]{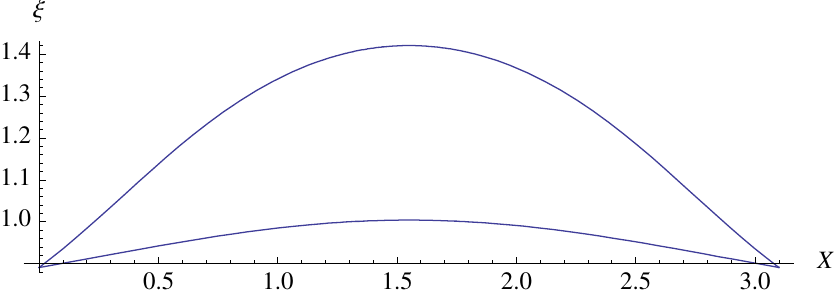} \\
				{$\Delta X = 3.10$} 
			\end{tabular}
			}} 
			\bigskip \\
			{\framebox{
			\begin{tabular}
				{c} 
				\includegraphics[width=1.6in]{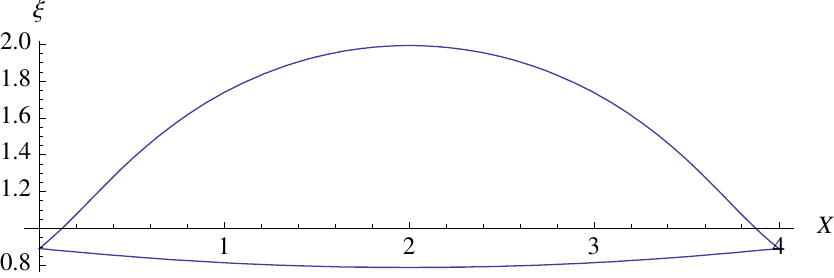} \\
				{$\Delta X = 4.00$} 
			\end{tabular}
			}} & {\framebox{
			\begin{tabular}
				{c} 
				\includegraphics[width=1.6in]{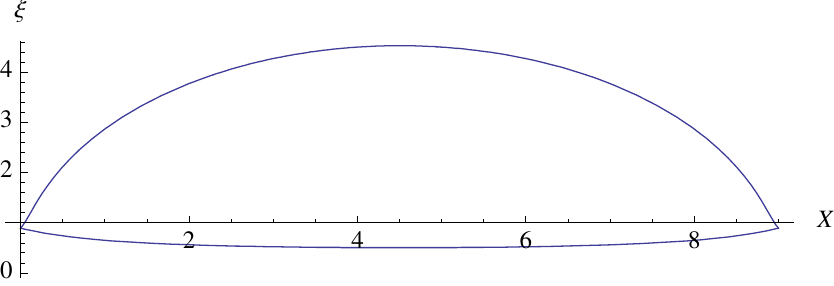} \\
				{$\Delta X = 9.00$} 
			\end{tabular}
			}} \\
		\end{tabular}
	\end{center}
	\caption{Geodesics for a sequence of displacements}\label{fig:dispplots} 
\end{figure}

For all displacements above a minimum value $\Delta X_{\text{min}} \simeq 2.95029$, we find the existence of two different geodesics. However, at a critical displacement $\Delta X_{cr} = \frac{\ell_{cr}}{P_m}$ the pair of geodesics flip from both being of the IR type to one of each type; this curious behavior is shown in Figure \ref{fig:dispplots}.

Perhaps more curious than the geodesic flipping is the existence of the minimum. For displacements smaller than this value, there are no UV or IR geodesics; the only geodesic is the path $\xi = \xi_0$, which is in fact a solution to the geodesic equations. This path corresponds to the critical momentum $P_{m}$.  Since it is limited to the critical momentum, this issue does not necessarily arise in a momentum-space discussion; however, a position-space argument begs for treatment of the horizontal path. Since our purpose is holography, a path fully in the holographic screen is questionable. It is, however, really the dominant path for small displacements. 

To show this, we find numerical solutions to the Hamilton-Jacobi equation 
\begin{equation}
	\label{eq:hjequation} g^{\mu \nu} 
	\partial_\mu S 
	\partial_\nu S = m^2 
\end{equation}
which governs the action $S = \int m\ ds$. (\ref{eq:hjequation}) is difficult to solve numerically, since
geodesics are singular paths of this partial differential equation so may not be used to impose
boundary values.

\begin{figure}
	\begin{center}
		\begin{tabular}{ccc}
		\includegraphics[width=2in]{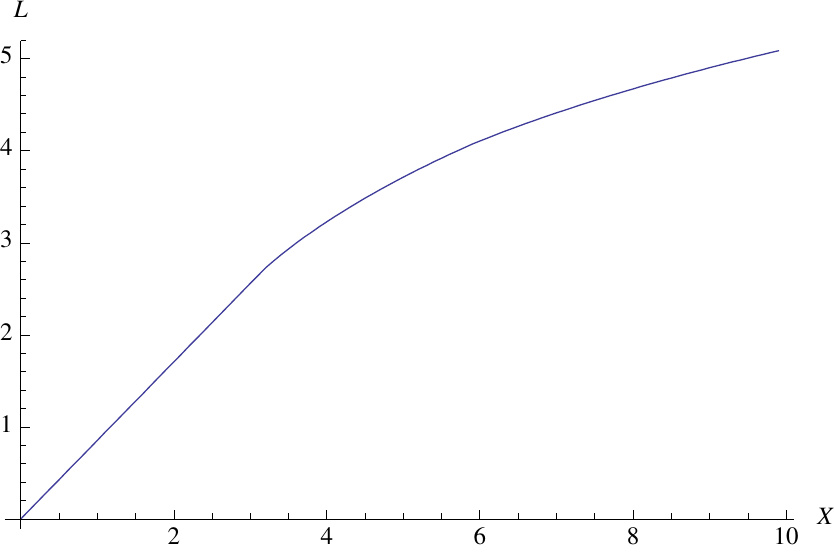} & \hspace{1in} &
		\includegraphics[width=2in]{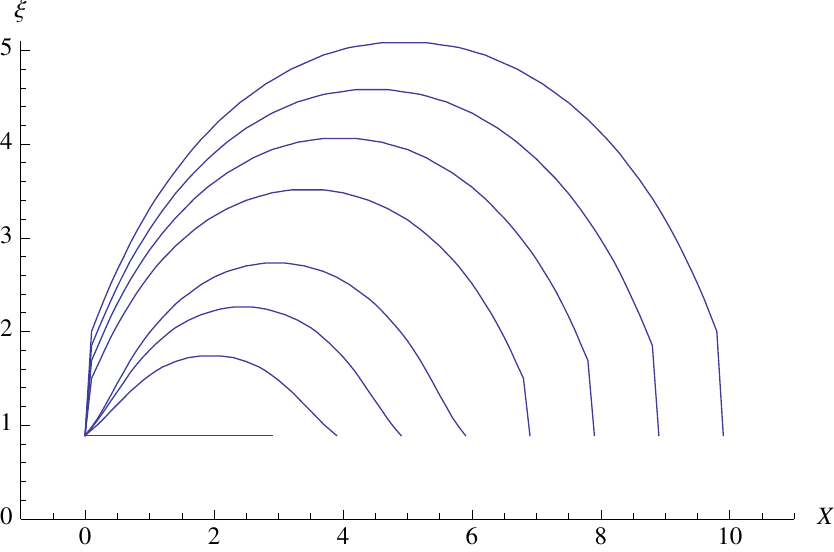} \\
		(a) & \hspace{1in} & (b)
		\end{tabular}
	\end{center}
	\caption{(a) Plot of geodesic lengths from Dijkstra's algorithm along $\xi = \xi_0$; (b) Minimum length paths from Dijkstra's algorithm}\label{fig:dijkaxislength} 
\end{figure}

We circumvent this difficulty by solving the closely related shortest path problem. The shortest path function must be a solution to (\ref{eq:hjequation}), and in fact may be considered \emph{the} solution of interest as argued in the Discussion section. Placing a lattice on the manifold and then using Dijkstra's shortest-path algorithm, we obtain a numerical solution for the action. Figure \ref{fig:dijkaxislength}(a) plots the action between two points on the boundary, and Figure \ref{fig:dijkaxislength}(b) shows the paths themselves. We see a sharp transition between the horizontal paths and the curved paths, which corresponds to a change between a linear displacement/length relation and a concave relation.
Based on these results, the position-space correlation function would be exponentially decaying for small distances, as opposed to the power-law behavior found in a local field theory.

\providecommand{\href}[2]{#2}\begingroup\raggedright\endgroup


\begin{thebibliography}{10}

\bibitem{Seiberg:2000ms}
N.~Seiberg, L.~Susskind, and N.~Toumbas, ``Strings in background electric
  field, space/time noncommutativity and a new noncritical string theory,''
  {\em JHEP} {\bf 06} (2000) 021,
  \href{http://xxx.lanl.gov/abs/hep-th/0005040}{{\tt hep-th/0005040}}.

\bibitem{Hashimoto:2000ys}
A.~Hashimoto and N.~Itzhaki, ``Traveling faster than the speed of light in
  non-commutative geometry,'' {\em Phys. Rev.} {\bf D63} (2001) 126004,
  \href{http://xxx.lanl.gov/abs/hep-th/0012093}{{\tt hep-th/0012093}}.

\bibitem{Seiberg:1999vs}
N.~Seiberg and E.~Witten, ``String theory and noncommutative geometry,'' {\em
  JHEP} {\bf 09} (1999) 032, \href{http://xxx.lanl.gov/abs/hep-th/9908142}{{\tt
  hep-th/9908142}}.

\bibitem{Bigatti:1999iz}
D.~Bigatti and L.~Susskind, ``Magnetic fields, branes and noncommutative
  geometry,'' {\em Phys. Rev.} {\bf D62} (2000) 066004,
  \href{http://xxx.lanl.gov/abs/hep-th/9908056}{{\tt hep-th/9908056}}.

\bibitem{Ganor:2006ub}
O.~J. Ganor, ``A new Lorentz violating nonlocal field theory from string-
  theory,'' {\em Phys. Rev.} {\bf D75} (2007) 025002,
  \href{http://xxx.lanl.gov/abs/hep-th/0609107}{{\tt hep-th/0609107}}.

\bibitem{Ganor:2007qh}
O.~J. Ganor, A.~Hashimoto, S.~Jue, B.~S. Kim, and A.~Ndirango, ``Aspects of
  Puff field theory,'' \href{http://xxx.lanl.gov/abs/hep-th/0702030}{{\tt
  hep-th/0702030}}.

\bibitem{Maldacena:1999mh}
J.~M. Maldacena and J.~G. Russo, ``Large N limit of non-commutative gauge
  theories,'' {\em JHEP} {\bf 09} (1999) 025,
  \href{http://xxx.lanl.gov/abs/hep-th/9908134}{{\tt hep-th/9908134}}.

\bibitem{Hashimoto:1999ut}
A.~Hashimoto and N.~Itzhaki, ``Non-commutative Yang-Mills and the AdS/CFT
  correspondence,'' {\em Phys. Lett.} {\bf B465} (1999) 142--147,
  \href{http://xxx.lanl.gov/abs/hep-th/9907166}{{\tt hep-th/9907166}}.

\bibitem{Hashimoto:1999yj}
A.~Hashimoto and N.~Itzhaki, ``On the hierarchy between non-commutative and
  ordinary supersymmetric Yang-Mills,'' {\em JHEP} {\bf 12} (1999) 007,
  \href{http://xxx.lanl.gov/abs/hep-th/9911057}{{\tt hep-th/9911057}}.

\bibitem{Cai:1999aw}
  R.~G.~Cai and N.~Ohta,
  ``On the thermodynamics of large N non-commutative super Yang-Mills
  theory,''
  Phys.\ Rev.\  D {\bf 61} (2000) 124012,
  \href{http://xxx.lanl.gov/abs/hep-th/9910092}{{\tt hep-th/9910092}}.

\bibitem{Klebanov:2000pp}
I.~R. Klebanov and J.~M. Maldacena, ``1+1 dimensional NCOS and its U(N) gauge
  theory dual,'' {\em Int. J. Mod. Phys.} {\bf A16} (2001) 922--935,
  \href{http://xxx.lanl.gov/abs/hep-th/0006085}{{\tt hep-th/0006085}}.

\bibitem{Bergman:2000cw}
A.~Bergman and O.~J. Ganor, ``Dipoles, twists and noncommutative gauge
  theory,'' {\em JHEP} {\bf 10} (2000) 018,
  \href{http://xxx.lanl.gov/abs/hep-th/0008030}{{\tt hep-th/0008030}}.

\bibitem{Maldacena:1997re}
J.~M. Maldacena, ``The large N limit of superconformal field theories and
  supergravity,'' {\em Adv. Theor. Math. Phys.} {\bf 2} (1998) 231--252,
  \href{http://xxx.lanl.gov/abs/hep-th/9711200}{{\tt hep-th/9711200}}.

\bibitem{Witten:1998qj}
E.~Witten, ``Anti-de Sitter space and holography,'' {\em Adv. Theor. Math.
  Phys.} {\bf 2} (1998) 253--291,
  \href{http://xxx.lanl.gov/abs/hep-th/9802150}{{\tt hep-th/9802150}}.

\bibitem{Bousso:1999xy}
R.~Bousso, ``A covariant entropy conjecture,'' {\em JHEP} {\bf 07} (1999) 004,
  \href{http://xxx.lanl.gov/abs/hep-th/9905177}{{\tt hep-th/9905177}}.

\bibitem{Ryu:2006bv}
S.~Ryu and T.~Takayanagi, ``Holographic derivation of entanglement entropy from
  AdS/CFT,'' {\em Phys. Rev. Lett.} {\bf 96} (2006) 181602,
  \href{http://xxx.lanl.gov/abs/hep-th/0603001}{{\tt hep-th/0603001}}.

\bibitem{Hubeny:2007xt}
V.~E. Hubeny, M.~Rangamani, and T.~Takayanagi, ``A covariant holographic
  entanglement entropy proposal,''
  \href{http://xxx.lanl.gov/abs/arXiv:0705.0016 [hep-th]}{{\tt arXiv:0705.0016
  [hep-th]}}.

\bibitem{Hawking:2000da}
S.~Hawking, J.~M. Maldacena, and A.~Strominger, ``deSitter entropy, quantum
  entanglement and AdS/CFT,'' {\em JHEP} {\bf 05} (2001) 001,
  \href{http://xxx.lanl.gov/abs/hep-th/0002145}{{\tt hep-th/0002145}}.

\bibitem{Maldacena:2001kr}
J.~M. Maldacena, ``Eternal black holes in anti-de-Sitter,'' {\em JHEP} {\bf 04}
  (2003) 021, \href{http://xxx.lanl.gov/abs/hep-th/0106112}{{\tt
  hep-th/0106112}}.

\bibitem{Brustein:2005vx}
R.~Brustein, M.~B. Einhorn, and A.~Yarom, ``Entanglement interpretation of
  black hole entropy in string theory,'' {\em JHEP} {\bf 01} (2006) 098,
  \href{http://xxx.lanl.gov/abs/hep-th/0508217}{{\tt hep-th/0508217}}.

\bibitem{Buniy:2005au}
R.~V. Buniy and S.~D.~H. Hsu, ``Entanglement entropy, black holes and
  holography,'' {\em Phys. Lett.} {\bf B644} (2007) 72--76,
  \href{http://xxx.lanl.gov/abs/hep-th/0510021}{{\tt hep-th/0510021}}.

\bibitem{Peet:1998wn}
A.~W. Peet and J.~Polchinski, ``UV/IR relations in AdS dynamics,'' {\em Phys.
  Rev.} {\bf D59} (1999) 065011,
  \href{http://xxx.lanl.gov/abs/hep-th/9809022}{{\tt hep-th/9809022}}.

\bibitem{Balasubramanian:1998sn}
V.~Balasubramanian, P.~Kraus, and A.~E. Lawrence, ``Bulk vs. boundary dynamics
  in anti-de Sitter spacetime,'' {\em Phys. Rev.} {\bf D59} (1999) 046003,
  \href{http://xxx.lanl.gov/abs/hep-th/9805171}{{\tt hep-th/9805171}}.

\bibitem{Balasubramanian:1998de}
V.~Balasubramanian, P.~Kraus, A.~E. Lawrence, and S.~P. Trivedi, ``Holographic
  probes of anti-de Sitter space-times,'' {\em Phys. Rev.} {\bf D59} (1999)
  104021, \href{http://xxx.lanl.gov/abs/hep-th/9808017}{{\tt hep-th/9808017}}.

\bibitem{Rozali:2000np}
M.~Rozali and M.~Van~Raamsdonk, ``Gauge invariant correlators in
  non-commutative gauge theory,'' {\em Nucl. Phys.} {\bf B608} (2001) 103--124,
  \href{http://xxx.lanl.gov/abs/hep-th/0012065}{{\tt hep-th/0012065}}.

\bibitem{Gross:2000ba}
D.~J. Gross, A.~Hashimoto, and N.~Itzhaki, ``Observables of non-commutative
  gauge theories,'' {\em Adv. Theor. Math. Phys.} {\bf 4} (2000) 893--928,
  \href{http://xxx.lanl.gov/abs/hep-th/0008075}{{\tt hep-th/0008075}}.

\bibitem{Martin:2003kp}
J.~Martin and R.~Brandenberger, ``On the dependence of the spectra of
  fluctuations in inflationary cosmology on trans-Planckian physics,'' {\em
  Phys. Rev.} {\bf D68} (2003) 063513,
  \href{http://xxx.lanl.gov/abs/hep-th/0305161}{{\tt hep-th/0305161}}.

\bibitem{Martin:2004um}
J.~Martin, ``Inflationary cosmological perturbations of quantum mechanical
  origin,'' {\em Lect. Notes Phys.} {\bf 669} (2005) 199--244,
  \href{http://xxx.lanl.gov/abs/hep-th/0406011}{{\tt hep-th/0406011}}.

\bibitem{Martin:2007bw}
J.~Martin, ``Inflationary perturbations: The cosmological Schwinger effect,''
  \href{http://xxx.lanl.gov/abs/arXiv:0704.3540 [hep-th]}{{\tt arXiv:0704.3540
  [hep-th]}}.

\bibitem{Kaloper:2002uj}
N.~Kaloper, M.~Kleban, A.~E. Lawrence, and S.~Shenker, ``Signatures of short
  distance physics in the Cosmic Microwave Background,'' {\em Phys. Rev.} {\bf
  D66} (2002) 123510, \href{http://xxx.lanl.gov/abs/hep-th/0201158}{{\tt
  hep-th/0201158}}.

\bibitem{Borunda:2006fx}
M.~Borunda and L.~Boubekeur, ``The effect of alpha' corrections in string gas
  cosmology,'' {\em JCAP} {\bf 0610} (2006) 002,
  \href{http://xxx.lanl.gov/abs/hep-th/0604085}{{\tt hep-th/0604085}}.

\bibitem{Holman:2006ny}
R.~Holman, L.~Mersini-Houghton, and T.~Takahashi, ``Cosmological avatars of the
  landscape. II: CMB and lss signatures,''
  \href{http://xxx.lanl.gov/abs/hep-th/0612142}{{\tt hep-th/0612142}}.

\bibitem{Chowdhury:2006pk}
B.~D. Chowdhury and S.~D. Mathur, ``Fractional brane state in the early
  universe,'' {\em Class. Quant. Grav.} {\bf 24} (2007) 2689--2720,
  \href{http://xxx.lanl.gov/abs/hep-th/0611330}{{\tt hep-th/0611330}}.

\bibitem{Spalinski:2007dv}
M.~Spalinski, ``On power law inflation in DBI models,'' {\em JCAP} {\bf 0705}
  (2007) 017, \href{http://xxx.lanl.gov/abs/hep-th/0702196}{{\tt
  hep-th/0702196}}.

\bibitem{Barnaby:2007yb}
N.~Barnaby and J.~M. Cline, ``Large nongaussianity from nonlocal inflation,''
  \href{http://xxx.lanl.gov/abs/arXiv:0704.3426 [hep-th]}{{\tt arXiv:0704.3426
  [hep-th]}}.

\bibitem{Kallosh:2007ig}
R.~Kallosh, ``On inflation in string theory,''
  \href{http://xxx.lanl.gov/abs/hep-th/0702059}{{\tt hep-th/0702059}}.

\bibitem{Brandenberger:2007bt}
R.~H. Brandenberger, ``String gas cosmology and structure formation: A brief
  review,'' \href{http://xxx.lanl.gov/abs/hep-th/0702001}{{\tt
  hep-th/0702001}}.

\bibitem{Tsujikawa:2003gh}
  S.~Tsujikawa, R.~Maartens and R.~Brandenberger,
  ``Non-commutative inflation and the CMB,''
  Phys.\ Lett.\  B {\bf 574} (2003) 141,
  \href{http://xxx.lanl.gov/abs/astro-ph/0308169}{{\tt astro-ph/0308169}}.

\bibitem{Koh:2007rx}
S.~Koh and R.~H. Brandenberger, ``Cosmological perturbations in non-commutative
  inflation,'' {\em JCAP} {\bf 0706} (2007) 021,
  \href{http://xxx.lanl.gov/abs/hep-th/0702217}{{\tt hep-th/0702217}}.

\bibitem{Brandenberger:2007rg}
R.~H. Brandenberger, ``String theory, space-time non-commutativity and
  structure formation,'' \href{http://xxx.lanl.gov/abs/hep-th/0703173}{{\tt
  hep-th/0703173}}.

\bibitem{Mukhanov:1990me}
V.~F. Mukhanov, H.~A. Feldman, and R.~H. Brandenberger, ``Theory of
  cosmological perturbations. part 1. classical perturbations. part 2. quantum
  theory of perturbations. part 3. extensions,'' {\em Phys. Rept.} {\bf 215}
  (1992) 203--333.

\bibitem{Mukhanov:2003xw}
V.~F. Mukhanov, ``CMB, quantum fluctuations and the predictive power of
  inflation,'' \href{http://xxx.lanl.gov/abs/astro-ph/0303077}{{\tt
  astro-ph/0303077}}.

\bibitem{Mukhanov:2005sc}
V.~Mukhanov, ``Physical foundations of cosmology,''. Cambridge, UK: Univ. Pr.
  (2005) 421 p.

\bibitem{Danielsson:2002kx}
U.~H. Danielsson, ``A note on inflation and transplanckian physics,'' {\em
  Phys. Rev.} {\bf D66} (2002) 023511,
  \href{http://xxx.lanl.gov/abs/hep-th/0203198}{{\tt hep-th/0203198}}.

\bibitem{Zamolodchikov:1986gt}
A.~B. Zamolodchikov, ``Irreversibility of the flux of the renormalization group
  in a 2d field theory,'' {\em JETP Lett.} {\bf 43} (1986) 730--732.

\bibitem{Sahakian:1999bd}
V.~Sahakian, ``Holography, a covariant c-function and the geometry of the
  renormalization group,'' {\em Phys. Rev.} {\bf D62} (2000) 126011,
  \href{http://xxx.lanl.gov/abs/hep-th/9910099}{{\tt hep-th/9910099}}.

\bibitem{Harmark:2000wv}
T.~Harmark, ``Supergravity and space-time non-commutative open string theory,''
  {\em JHEP} {\bf 07} (2000) 043,
  \href{http://xxx.lanl.gov/abs/hep-th/0006023}{{\tt hep-th/0006023}}.

\bibitem{Sahakian:2001xz}
V.~Sahakian, ``The large M limit of non-commutative open strings at strong
  coupling,'' {\em Nucl. Phys.} {\bf B621} (2002) 62--100,
  \href{http://xxx.lanl.gov/abs/hep-th/0107180}{{\tt hep-th/0107180}}.

\end{thebibliography}
\end{document}